\documentclass[twocolumn,twocolappendix]{aastex7}
\usepackage{color}
\usepackage{graphicx}
\usepackage{amsmath}    
\usepackage{amssymb}
\usepackage{mathrsfs}
\usepackage{times}
\usepackage{xcolor}
\usepackage{xspace}
\usepackage{datetime}
\usepackage{version}    
\usepackage{placeins}
\usepackage[export]{adjustbox}
\usepackage{natbib}

\newcommand{\simgt}{\lower.5ex\hbox{$\; \buildrel > \over \sim \;$}}
\newcommand{\simlt}{\lower.5ex\hbox{$\; \buildrel < \over \sim \;$}}

\newcommand{\clumiplus}{\textsc{clumi+}\xspace}
\newcommand{\clumi}{\textsc{clumi}\xspace}
\newcommand{\percent}{\ensuremath{\%}\xspace}
\newcommand{\Om}{\Omega_\mathrm{m}}
\newcommand{\OL}{\Omega_\Lambda}
\newcommand{\Ob}{\Omega_\mathrm{b}}
\newcommand{\zs}{z_s}
\newcommand{\zl}{z_l}
\newcommand{\Map}{M_\mathrm{ap}}
\newcommand{\vesc}{v_\mathrm{esc}}
\newcommand{\SigmaCrit}{\Sigma_\mathrm{cr}}
\newcommand{\SigmaCritInf}{\Sigma_{\mathrm{cr},\infty}}
\newcommand{\Rsp}{R_\mathrm{sp}}
\newcommand{\Rinf}{R_\mathrm{inf}}
\newcommand{\Rein}{\theta_\mathrm{Ein}}
\newcommand{\mcut}{m_\mathrm{cut}}
\newcommand{\NWL}{N_\mathrm{WL}}
\newcommand{\NSL}{N_\mathrm{SL}}
\newcommand{\NEV}{N_\mathrm{esc}}
\newcommand{\thetamin}{\theta_\mathrm{min}}
\newcommand{\thetamax}{\theta_\mathrm{max}}
\newcommand{\kpch}{\,h^{-1}\,\mathrm{kpc}}
\newcommand{\Mpch}{\,h^{-1}\,\mathrm{Mpc}}
\newcommand{\Mpc}{\,\mathrm{Mpc}}
\newcommand{\Msunh}{\,h^{-1}\,M_\odot}
\newcommand{\Msun}{\,M_\odot}
\newcommand{\LCDM}{$\Lambda$CDM\xspace}
\newcommand{\Planck}{Planck\xspace}
\newcommand{\HST}{HST\xspace}
\newcommand{\Euclid}{Euclid\xspace}
\newcommand{\ROSAT}{ROSAT\xspace}
\newcommand{\FIG}{./}

\def\btheta{\mbox{\boldmath $\theta$}} 
\def\bnabla{\mbox{\boldmath $\nabla$}}
\def\bs{\mbox{\boldmath $s$}}
\def\bc{\mbox{\boldmath $c$}}
\def\bm{\mbox{\boldmath $m$}}
\def\bd{\mbox{\boldmath $d$}}
\def\bp{\mbox{\boldmath $p$}}
\def\br{\mbox{\boldmath $r$}}

\def\ge{\geqslant}
\def\le{\leqslant}

\shorttitle{Combining Dynamics and Weak Lensing around Galaxy Clusters}
\shortauthors{Umetsu et~al.}

\begin{document}
\title{Cluster Lensing Mass Inversion (CLUMI+): Combining Dynamics and Weak Lensing around Galaxy Clusters}

\correspondingauthor{Keiichi Umetsu}

\author[0000-0002-7196-4822]{Keiichi Umetsu}
\affiliation{Academia Sinica Institute of Astronomy and Astrophysics (ASIAA), No.~1, Section~4, Roosevelt Road, Taipei 106216, Taiwan} 
\email{keiichi@asiaa.sinica.edu.tw}

\author[0000-0003-0981-330X]{Michele Pizzardo}
\affiliation{Academia Sinica Institute of Astronomy and Astrophysics (ASIAA), No.~1, Section~4, Roosevelt Road, Taipei 106216, Taiwan} 
\affiliation{Department of Astronomy and Physics, Saint Mary's University, 923 Robie Street, Halifax, NS-B3H3C3, Canada}
\email{michele.pizzardo@gmail.com} 

\author[0000-0002-4986-063X]{Antonaldo Diaferio}
\affiliation{Dipartimento di Fisica, Universit\`a di Torino, via P. Giuria 1, I-10125 Torino, Italy} 
\affiliation{Istituto Nazionale di Fisica Nucleare (INFN), Sezione di Torino, via P. Giuria 1,  I-10125 Torino, Italy}
\email{diaferio@ph.unito.it}

\author[0000-0002-9146-4876]{Margaret J. Geller}
\affiliation{Center for Astrophysics--Harvard \& Smithsonian, 60 Garden Street, Cambridge, MA 02138, USA}
\email{mgeller@cfa.harvard.edu}

\begin{abstract}
We present \clumiplus, a self-consistent, multiprobe methodology for reconstructing the mass distribution in and around galaxy clusters by combining gravitational lensing and dynamical observations. Building on the joint likelihood framework of \citet{Umetsu2013}, \clumiplus integrates weak-lensing shear and magnification data with projected escape velocity measurements in the cluster infall region, yielding tighter constraints on the gravitational potential without relying on equilibrium assumptions. The mass distribution is modeled using a flexible, piecewise-defined convergence profile that characterizes the azimuthally averaged surface mass density within the lensing field, transitioning to a projected power-law form at larger radii 
where phase-space constraints complement lensing.
Additional strong-lensing constraints are incorporated via central aperture-mass measurements, enabling full-scale mass reconstruction from the cluster core to the outskirts. We validate \clumiplus using synthetic weak-lensing and phase-space data for a massive cluster from the IllustrisTNG simulations, demonstrating unbiased recovery of projected and three-dimensional mass profiles and achieving 10\percent--30\percent improvement in precision at large radii.
As a case study, we apply \clumiplus to A2261, combining Subaru and Hubble Space Telescope weak+strong lensing data with spectroscopic measurements from the Hectospec Cluster Survey. This analysis demonstrates the power of multiprobe, equilibrium-free modeling for robust cluster mass reconstruction.
\end{abstract}
 
\keywords{
\uat{Dark matter distribution}{356};
\uat{Galaxy clusters}{584};
\uat{Gravitational fields}{667};
\uat{Weak gravitational lensing}{1797};
\uat{Strong gravitational lensing}{1643};
\uat{Redshift surveys}{1378};
\uat{Large-scale structure of the universe}{902}
}

\section{Introduction}
\label{sec:intro}

Galaxy clusters represent the largest self-gravitating systems in the universe, assembled through hierarchical merging governed by the growth of cosmic structure. As such, clusters play a central role in testing models of background cosmology and structure formation \citep[][]{Allen2011rev,Pratt2019}. A key requirement for cluster-based cosmological studies, such as cluster abundance measurements \citep{Rosati+2002}, is the precise characterization of the dark matter distribution in and around cluster halos, which encodes critical information about the physics of structure growth. In this context, the standard $\Lambda$ cold dark matter (\LCDM) paradigm and its viable extensions \citep[][]{Spergel+Steinhardt2000,Schive2014psiDM} predict distinct, testable signatures in the density and velocity structure of galaxy clusters \citep{Taylor+Navarro2001,Hjorth+2010DARKexp,Han2023,Williams2025}.

Gravitational lensing provides a direct and robust probe of the mass distribution in clusters, which is dominated by dark matter, without relying on assumptions about the dynamical state of the system. Strong lensing produces highly distorted and multiply imaged background galaxies in the dense cores of massive halos \citep{Kneib+Natarajan2011, Meneghetti2021}, whereas weak lensing maps the projected mass distribution on larger scales by measuring coherent distortions and magnification effects in background galaxy populations \citep{Bartelmann+Schneider2001,Umetsu2020rev}. 

A key advantage of weak lensing is that the observed signal can be directly interpreted in a free-form way across a wide range of spatial scales beyond the cluster core, enabling both radial and two-dimensional (2D) reconstructions of the projected mass distribution \citep[][]{Jee+2005,UB2008,Bradac2006,Jauzac+2012}. Accordingly, weak lensing has been established as a critical tool for calibrating observable--mass relations in cluster samples selected through X-ray, Sunyaev--Zel'dovich (SZ) effect, or optical surveys \citep{WtG3,Hoekstra2015CCCP,Okabe+Smith2016,Sereno2017psz2lens,Bocquet2019,Umetsu2020xxl,Chiu2022,Grandis2024}.

Spectroscopic data offer an independent and complementary avenue for probing cluster mass distributions by characterizing the velocity structure around clusters in projected phase space. Redshift diagrams constructed from spectroscopic measurements of tracer galaxies enable dynamical mass estimates through a variety of techniques. In particular, Jeans modeling and virial analyses infer cluster masses from the observed line-of-sight velocity dispersion of member galaxies \citep[][]{Mamon2013,Biviano2013,Pizzuti2016clash,Sereno2025}, under assumptions of dynamical equilibrium and orbital velocity anisotropy. In contrast, the caustic technique identifies the outer envelope of the galaxy distribution in redshift space \citep{Diaferio+Geller1997,Diaferio1999,Serra2011}, which traces the projected escape velocity profile and enables mass estimates extending into the infall regions \citep{Geller1999,Geller2013,Rines2013HeCS}. Joint analyses combining lensing and dynamical tracers have also been employed to test models of modified gravity in cluster outskirts \citep{Pizzuti2017clash,Pizzuti2021,Butt2024,Butt2025} and to constrain the orbital anisotropy of galaxy populations within clusters \citep[][]{Lemze+2009,Stark2019}.

In this paper, we develop Cluster Lensing Unbiased Multiprobe Inversion with spectroscopic constraints (\clumiplus), an extended multiprobe mass reconstruction framework that builds on the \clumi algorithm introduced by \citet{Umetsu+2011} and \citet{Umetsu2013} (see also \citealt{Umetsu2015} and \citealt{Umetsu2018clump3d} for a 2D implementation). The original framework combines weak-lensing shear and magnification profiles with strong-lensing aperture-mass constraints, enabling flexible modeling of cluster surface mass density profiles using a piecewise-defined approach, with minimal prior assumptions. \clumiplus expands this formalism by incorporating projected escape velocity measurements derived from redshift diagrams. It is especially powerful in the cluster outskirts and infall regions, where spectroscopic phase-space information provides complementary constraints on the three-dimensional (3D) mass distribution in an equilibrium-free manner. This extension enhances reconstruction precision and facilitates physical interpretation of dynamical structure, including the dynamical boundary of cluster halos \citep[][]{Diemer+Kravtsov2014,Pizzardo2024splash}.

This paper is organized as follows: In Section~\ref{sec:lens}, we review the theoretical foundations of cluster--galaxy lensing. Section~\ref{sec:caustic} introduces the projected escape velocity formalism and the caustic technique. In Section~\ref{sec:clumiplus}, we present the \clumiplus methodology, describing our multiprobe mass modeling framework and its Bayesian implementation. In Section~\ref{sec:tng}, we validate \clumiplus using synthetic observations of a massive simulated cluster from the IllustrisTNG suite. In Section~\ref{sec:a2261}, we apply \clumiplus to real multiprobe observations of the galaxy cluster A2261. Finally, we summarize our findings and conclude in Section~\ref{sec:summary}.

Throughout this paper, we adopt a spatially flat $\Lambda$CDM cosmology based on the \Planck 2015 results \citep{ade2016planck}, with $\Om = 0.3089$, $\OL = 0.6911$, and a Hubble constant of $H_0 = 100\,h$~km\,s$^{-1}$\,Mpc$^{-1}$ with $h = 0.6774$. Projected and 3D cluster-centric radii are denoted as $r_\perp$ and $r$, respectively. Spherical overdensity masses are defined with respect to either the critical density or mean background density of the universe at the cluster redshift, with $M_{\Delta\mathrm{c}}$ or $M_{\Delta\mathrm{m}}$ denoting the mass enclosed within a radius where the mean overdensity equals $\Delta_\mathrm{c}$ or $\Delta_\mathrm{m}$, respectively. In this paper, we use overdensity values of $\Delta_\mathrm{c} = 200,\ 500$ and $\Delta_\mathrm{m} = 200$. All quoted uncertainties are at the $1\sigma$ confidence level unless otherwise stated.

\section{Cluster--Galaxy Lensing Theory}
\label{sec:lens}

The effects of weak gravitational lensing on background galaxies are characterized by the convergence, $\kappa$ (the isotropic magnification term), and the complex shear with spin-2 rotational symmetry, $\gamma=|\gamma|e^{2i\phi_\gamma}$, in 2D projection. For a comprehensive review of cluster galaxy weak lensing, see \citet{Umetsu2020rev}, and for strong lensing, refer to \citet{Meneghetti2021} and \citet{Natarajan2024}.

\subsection{Basics of Cluster--galaxy Weak Lensing}
\label{subsec:basics_lens}

The lensing convergence is defined as $\kappa=\Sigma/\SigmaCrit$, where $\Sigma$ is the surface mass overdensity of a lens in units of the critical surface density for gravitational lensing:
\begin{equation}
 \SigmaCrit(\zl, \zs) = \frac{c^2}{4\pi G} \frac{D_s}{D_l D_{ls}} = \frac{c^2}{4\pi G D_l \beta},
\end{equation}
with $c$ the speed of light, $G$ the gravitational constant, and $D_l(\zl)$, $D_s(\zs)$, and $D_{ls}(\zl,\zs)$ the observer--lens, observer--source, and lens--source angular diameter distances, respectively. The geometric lensing efficiency, $\beta(\zl,\zs)=D_{ls}/D_s$, determines the strength of lensing as a function of the lens redshift $\zl$ and source redshift $\zs$, where we set $\beta(\zl,\zs) = 0$ for unlensed sources with $\zs \le \zl$. The shear and convergence thus depend on both $(\zl, \zs)$ and the image position $\btheta = \br_\perp/D_l$. For conciseness, we omit the explicit dependence on $\zl$ in the following discussion.

The gravitational shear $\gamma=\gamma_1+i\gamma_2$ is directly observable from the image ellipticities of background galaxies in the weak-lensing limit, $|\kappa|\ll 1$. The primary observable in weak lensing is the reduced shear $g=g_1+ig_2$,
\begin{equation}
\label{eq:redshear}
g =\frac{\gamma}{1-\kappa},
\end{equation}
which remains invariant under the global transformation $\kappa(\btheta)\to \lambda \kappa(\btheta) + 1-\lambda$ and $\gamma(\btheta)\to \lambda \gamma(\btheta)$ with an arbitrary constant $\lambda$ for a given source plane. The term $1-\lambda\equiv \kappa_0$ represents an additive constant convergence. This transformation, known as the mass-sheet degeneracy \citep[][]{Schneider+Seitz1995}, can be broken or alleviated, for example, by measuring the magnification factor in subcritical regions, outside the critical curves:
\begin{equation}
\label{eq:mu}
\mu = \frac{1}{(1-\kappa)^2-|\gamma|^2}.
\end{equation}
The magnification factor $\mu$ transforms as $\mu(\btheta)\to \lambda^{-2}\mu(\btheta)$.

The reduced shear $g_{1,2}$ can be decomposed into the tangential component $g_+=\gamma_+/(1-\kappa)$ and the $45^\circ$-rotated cross-shear component $g_\times=\gamma_\times/(1-\kappa)$ with respect to a given reference point. The tangential- and cross-shear components ($\gamma_+,\gamma_\times$) averaged around a circle of radius $\theta$ satisfy the following identities for an arbitrary choice of the center:
\begin{equation}
\label{eq:gammatx}
  \begin{aligned}
  \gamma_+(\theta) &= \overline{\kappa}(<\theta)-\kappa(\theta) \equiv 
   \frac{\Delta\Sigma(r_\perp)}{\SigmaCrit},\\
   \gamma_\times(\theta) &= 0,
  \end{aligned}
\end{equation}
where $\kappa(\theta)=\Sigma(r_\perp)/\SigmaCrit$ is the azimuthally averaged convergence at radius $\theta$ and $\overline{\kappa}(<\theta)=\overline{\Sigma}(<r_\perp)/\SigmaCrit$ is the average convergence interior to $\theta$.
In Equation (\ref{eq:gammatx}), we have introduced the excess surface mass density $\Delta\Sigma(r_\perp)=\overline{\Sigma}(<r_\perp)-\Sigma(r_\perp)$. 
In the absence of higher-order effects, the azimuthally averaged cross component of weak lensing is expected to vanish. Thus, the presence of $g_\times$ distortions serves as a null test for systematic errors.

\subsection{Source Redshift Distribution}
\label{subsec:Nz}

For weak-lensing analysis, we consider a population of source galaxies characterized by their intrinsic (unlensed) redshift distribution, $\overline{N}(\zs)$. 
Different selection criteria and quality cuts are typically applied for measuring shear and magnification effects, leading to distinct $\overline{N}(\zs)$ distributions for each effect.
We denote the redshift distribution functions for the shear and magnification measurements as $\overline{N}_g(z)$ and $\overline{N}_\mu(z)$, respectively.

We express source-averaged moments $\langle\beta^n\rangle_X$ ($n=1,2,\dots$) for a given population ($X=g,\mu$) as
\begin{equation}
\label{eq:depth}
\langle\beta^n\rangle_X =\left[
\int_0^\infty\!dz\, \overline{N}_X(z) \beta^n(\zl,z)\right]
\left[
\int_0^\infty\!dz\, \overline{N}_X(z)
\right]^{-1}.
\end{equation} 
In general, $\overline{N}(z)$ can include foreground galaxies. The contribution from unlensed objects with $\beta=0$ is thus accounted for in the calculation of $\langle\beta^n\rangle_X$.

We introduce the relative lensing efficiency of each source population:
\begin{equation}
 \langle W^n\rangle_X = \langle\beta^n\rangle_X  / \beta_\infty^n
\end{equation} 
with $\beta_\infty = \beta(\zl, z_{s,\infty})$ defined with respect to a reference source in the far background \citep{Bartelmann+Schneider2001}. We use a reference redshift of $z_{s,\infty}= 20,000$, as adopted in the CLASH program \citep{Umetsu2014clash,Merten2015clash}. The associated critical surface density is
\begin{equation}
 \SigmaCritInf(\zl)=\frac{c^2}{4\pi G D_l(\zl)} \frac{1}{\beta_{\infty}(\zl)}.
\end{equation}
Hereafter we use the far-background convergence $\kappa_\infty(\btheta)\equiv \Sigma(\br_\perp)/\SigmaCritInf$ to describe the projected mass distribution of the cluster lens.

\subsection{Weak Lensing: Tangential Distortion}
\label{subsec:gt}

From shape measurements of background galaxies, we construct azimuthally averaged radial profiles of the tangential distortion, $g_+(\theta)$, and the cross component, $g_\times(\theta)$, in subcritical regions as functions of cluster-centric angular radius $\theta$. The source-averaged expectation value for the reduced tangential shear, $\langle g_+ \rangle(\theta)$, can be approximated as follows \citep[][]{Umetsu2020rev}:
\begin{equation}
 \label{eq:gt}
 \begin{aligned}
  \langle g_+\rangle(\theta) &= \left[\int\!dz\,N_g(z) g_+(\theta,z) \right] 
\left[\int\!dz\,N_g(z) \right]^{-1}\\
&\simeq \frac{\langle \gamma_+\rangle(\theta)}{1-f_{g}\langle\kappa\rangle(\theta)},
 \end{aligned}
\end{equation}
where 
$\langle\kappa\rangle(\theta)=\langle W\rangle_g \kappa_\infty(\theta)$ and $\langle\gamma_+\rangle(\theta)=\langle W\rangle_g [\overline{\kappa}_\infty(<\theta)-\kappa_\infty(\theta)]$ are the source-averaged convergence and tangential shear, respectively, and $f_{g}=\langle W^2\rangle_g/\langle W\rangle_g^2$ is a dimensionless correction factor of the order unity.\footnote{Gravitational lensing observables generally depend nonlinearly on the lensing potential. Consequently, averaging over the source redshift distribution introduces nonlinear effects. In the weak-lensing regime, however, these effects are typically well approximated by low-order corrections involving the moments $\langle W^n \rangle$.}

\subsection{Weak Lensing: Magnification Bias}
\label{subsec:magbias}

In the presence of gravitational lensing, the source counts observed for a given magnitude cut $\mcut$ are modified depending on the intrinsic slope of the source luminosity function. For details, see Section~5 of \citet{Umetsu2020rev}.

\subsubsection{Density Depletion and Enhancement}
\label{subsubsec:magbias}

The magnification bias for a flux-limited sample of background galaxies is defined as the ratio of their magnified counts $n_\mu(\theta|<\mcut)$ around a cluster lens to the unlensed mean counts $\overline{n}_\mu(<\mcut)$ without lensing:
\begin{equation}
\label{eq:mag}
\begin{aligned} 
\langle b_\mu\rangle(\theta) = \frac{n_\mu(\theta|<\mcut)}{\overline{n}_\mu(<\mcut)}
=& \left[\int\!dz\,N_\mu(\theta,z|<\mcut)\right] \\
 &\times\left[\int\!dz\,\overline{N}_\mu(z|<\mcut)\right]^{-1},
\end{aligned}
\end{equation}
where $N_\mu(\theta,z)$ and $\overline{N}_\mu(z)$ are the lensed and unlensed redshift distribution functions of the flux-limited source sample, respectively (Section~\ref{subsec:Nz}).

The mass-sheet degeneracy can be lifted by combining shear and magnification measurements (Equations (\ref{eq:gt}) and (\ref{eq:mag})), provided that the unlensed source background density $\overline{n}_\mu(<\mcut)$ is known or can be estimated from data.

In weak lensing, the change in magnitude, $\delta m=2.5\log_{10}\mu$, due to magnification is small compared to the range of magnitudes over which the slope of the luminosity function varies. The source counts $\overline{N}_\mu(<\mcut)$ can thus be locally approximated by a power law at a given cutoff magnitude $\mcut$. Accordingly, we can interpret the magnification bias observed in subcritical regions by \citep{Umetsu2020rev}
\begin{equation}
\label{eq:magbias}
 \langle b_\mu\rangle(\theta)\simeq \langle\mu^{-1}\rangle^{1-\alpha}(\theta),
\end{equation}
where $\alpha$ is the logarithmic count slope evaluated at $\mcut$:\footnote{Instead of $\alpha$, a notation of $s\equiv 0.4\alpha$ is often used in the literature \citep[e.g.,][]{BTU+05}. The count slope $\alpha$ is related to the faint-end slope $\alpha_\mathrm{LF}$ of the source luminosity function as $\alpha=-(\alpha_\mathrm{LF}+1)$.}
\begin{equation}
 \label{eq:slope}
 \alpha(\mcut) = 2.5\frac{d\log_{10}\overline{N}_\mu(<m)}{dm}\Bigg|_{\mcut},
\end{equation}
The term $\langle \mu^{-1}\rangle$ represents the source-averaged inverse magnification,
which is approximated by \citep[][]{Umetsu2020rev}
\begin{equation}
\langle\mu^{-1}\rangle(\theta)
 = \left[1- \langle\kappa\rangle(\theta)\right]^2 - \langle\gamma_+\rangle^2(\theta),
\end{equation}
with $\langle\kappa\rangle(\theta)=\langle W\rangle_\mu \kappa_\infty(\theta)$
and  $\langle\gamma_+\rangle(\theta)=\langle W\rangle_\mu [\overline{\kappa}_\infty(<\theta)-\kappa_\infty(\theta)]$.

Equation (\ref{eq:magbias}) shows that a net count depletion ($\langle b_\mu\rangle < 1$) occurs when $\alpha < 1$, as the geometric area distortion dominates in magnification bias. In this depletion regime, a key advantage is that the effect is largely insensitive to the detailed form of the source luminosity function. For a maximally depleted population with $\alpha=0$, the bias factor simplifies to $\langle b_\mu\rangle=\langle\mu^{-1}\rangle$. Conversely, in the density enhancement regime ($\alpha > 1$), interpreting the observed lensing signal requires precise knowledge of the intrinsic luminosity function \citep{Chiu2020hscmag}, particularly in the nonlinear regime, where the amplification factor is large (e.g., $\mu \simgt 1.5$).

The net magnification effect on source counts is expected to vanish when $\alpha = 1$. Hence, lensing magnification provides a natural null test  \citep{Chiu2020hscmag,Umetsu2022}, enabling an assessment of residual biases that may be present in the lensing-selected sample. 

\subsubsection{Multipopulation Magnification Bias}
\label{subsubsec:relmagbias}

Deep multiband photometry across a broad wavelength range enables the selection of distinct background galaxy populations in color--color space \citep[][]{Medezinski+2010,Medezinski2018src}. Since a fixed flux limit corresponds to different intrinsic luminosities at different redshifts, the source counts of color-selected populations sample different regimes of magnification bias, from density enhancement to depletion. Incorporating independent count measurements from multiple source populations offers two key advantages for cluster lensing analyses. First, it improves the statistical precision of lensing constraints \citep{Umetsu2013}. Second, it reduces sensitivity to the intrinsic angular clustering of background galaxies, as distinct source populations are expected to be spatially uncorrelated \citep{Broadhurst1995}.

As demonstrated by \citet{Umetsu2013}, both \textsc{clumi} and \textsc{clumi}+ support magnification bias constraints from multiple background populations in a joint likelihood framework. In this study, however, we restrict our analysis to a single source population in the depletion regime for simplicity.

\subsection{Strong Lensing: Aperture Mass}
\label{subsec:sl}

Detailed modeling of the deflection field, $\bnabla \psi(\btheta)$, using sets of multiply lensed images with spectroscopic redshifts enables precise determination of the locations of critical curves. The lensing potential $\psi(\btheta)$ is given by  
\begin{equation}
\psi(\btheta) = \frac{1}{\pi} \int \ln(|\btheta - \btheta'|) \, \kappa(\btheta') \, d^2\theta',
\end{equation}  
where critical curves correspond to regions where the Jacobian determinant vanishes, $\det[\mathsf{A}(\btheta)] = 0$, and $\mathsf{A}(\btheta)$ is the Jacobian matrix of the lens mapping. The locations of these critical curves provide tight constraints on the projected mass enclosed within them.

In this context, the term ``Einstein radius'' commonly refers to the size of the outer critical curve \citep{Meneghetti2021}. The effective Einstein radius is defined as  
\begin{equation}
\Rein = \sqrt{A_\mathrm{cr}/\pi},
\end{equation}  
where $A_\mathrm{cr}$ is the area enclosed by the outer critical curve.  

The aperture mass enclosed within a cylinder of radius $\theta$ is defined as  
\begin{equation}
\label{eq:map_sl}
 \Map(<\theta) = \SigmaCritInf D_l^2 \int_{|\btheta'|<\theta}\! \kappa_\infty(\btheta') \, d^2\theta',
\end{equation}
When evaluated near the Einstein radius $\Rein$, $\Map(<\theta)$ becomes less sensitive to assumptions about the underlying mass model and is tightly constrained by strong-lensing observations \citep[][]{Umetsu2012,Gonzalez2021a,Gonzalez2021b}. This makes it a key observable in the strong-lensing regime \citep[][]{Coe+2010}.

As the Einstein radius $\Rein(z)$ increases with source redshift $z$, strongly lensed images from sources at different redshifts form nested configurations spanning a broader radial range. Consequently, multiple sets of strongly lensed images from different source redshifts can be used to constrain $\Map(<\theta)$ over an extended radial range \citep[][]{Coe2012, Umetsu2012, Umetsu2016clash}.

\section{Projected Escape Velocity Formalism}
\label{sec:caustic}

\subsection{Caustic Technique and Projected Phase Space}
\label{subsec:caustic}

Complementing gravitational lensing, redshift diagrams in projected phase space provide a powerful probe of the velocity structure around galaxy clusters. In this study, we focus on the identification and use of ``caustic'' boundaries, which delineate the velocity envelope of tracer galaxies in redshift space. These boundaries trace the projected escape velocity profile, which is directly related to the underlying 3D gravitational potential of the cluster.

The caustic technique \citep{Diaferio+Geller1997,Diaferio1999} identifies these escape velocity edges from the distribution of galaxies in projected phase space. 
In the caustic algorithm, spherical symmetry is assumed in the dynamical interpretation, specifically when relating spherically averaged velocity moments of tracer galaxies to the underlying gravitational potential \citep{Diaferio1999,Serra2011}. Under this assumption,
the resulting caustic amplitude, $\mathcal{A}(r_\perp) > 0$, is determined as a function of the projected cluster-centric radius $r_\perp$ and interpreted as a local estimate of the Newtonian escape velocity:\footnote{Escape velocity, $\vesc(r)$, is defined by the gravitational potential $\Phi_\mathrm{N}(r)$ at radius $r$, independent of the dynamical state or orbits of tracer galaxies. Even in the infall region, where tracers are outside the halo and on first approach, $\vesc(r)$ remains a well-defined upper bound for bound motion. In the caustic algorithm, this quantity reflects a kinematic threshold, not the binding status of the tracers.}
\begin{equation}
\label{eq:caustic}
 \mathcal{A}^2(r_\perp)  = 
\frac{\vesc^2(r_\perp)}{\mathcal{G}} =
\frac{-2\Phi_\mathrm{N}(r_\perp)}{\mathcal{G}},
\end{equation}
where $\vesc(r_\perp)$ denotes the 3D escape velocity evaluated at the spherical radius $r = r_\perp$, $\Phi_\mathrm{N}(r_\perp)$ is the Newtonian gravitational potential at the same radius,
and $\mathcal{G}$ is a dimensionless \textit{depletion factor} that quantifies the reduction of the observed velocity envelope relative to the true 3D escape velocity. The caustic technique has been tested and calibrated against cosmological simulations \citep{Diaferio1999,Serra2011,Pizzardo2023ct}, including dedicated tests on the TNG300-1 run of the IllustrisTNG simulations \citep{Pizzardo2023ct}.

Equation~(\ref{eq:caustic}) reflects the Newtonian interpretation of escape velocity adopted in \clumiplus, which neglects cosmic expansion and acceleration. While this approximation is not directly validated, the reliability of \clumiplus is assessed through its ability to recover unbiased mass profiles in both 2D and 3D, using synthetic observations of a realistic cluster from the IllustrisTNG simulation suite (Section~\ref{sec:tng}).

Throughout this study, we refer to $\mathcal{A}(r_\perp)$ as the \textit{projected escape velocity}, following the convention of the caustic technique \citep{Diaferio1999}. This quantity characterizes the amplitude of the observed velocity envelope in redshift space and is distinct from the true 3D escape velocity.

\subsection{Physical Interpretation of the Depletion Factor \texorpdfstring{$\mathcal{G}$}{G}}

In the caustic algorithm developed by \citet{Diaferio1999}, the depletion factor $\mathcal{G}$ is interpreted as a \textit{geometric} effect arising from the orbital velocity anisotropy of tracer galaxies. 
Under the assumption of negligible cluster rotation,\footnote{When the cluster rotation is negligible, $\langle v^2_\theta \rangle = \langle v^2_\phi\rangle = \langle v^2_\mathrm{los} \rangle$ and $\langle v_r^2 \rangle = \langle v^2 \rangle - 2\langle v^2_\mathrm{los} \rangle$, allowing the escape velocity anisotropy to be characterized via the line-of-sight velocity dispersion.}
it is related to the velocity anisotropy parameter $\beta_v(r)$ evaluated at $r=r_\perp$ by
\begin{equation}
\label{eq:gbeta}
  \mathcal{G} = \frac{\langle\vesc^2\rangle(r_\perp)}{\langle v^2_\mathrm{esc,los}\rangle(r_\perp)} = \frac{3-2\beta_v(r_\perp)}{1-\beta_v(r_\perp)},
\end{equation}
where the brackets $\langle\cdot\rangle(r)$ denote an average over tracer velocities within a spherical shell at 3D radius $r$, $v_\mathrm{esc,los}$ represents the line-of-sight component of the escape velocity, and $\beta_v$ is defined as \citep{Binney2008}:
\begin{equation}
  \beta_v(r) = 1-\frac{\langle v^2_\theta\rangle(r) + \langle v^2_\phi\rangle(r)}{2\langle v^2_r\rangle(r)},
\end{equation}
with $v_r$, $v_\theta$, and $v_\phi$ denoting the radial, polar, and azimuthal components, respectively, of the 3D velocity of a tracer galaxy. Theoretically, $\beta_v$ and $\mathcal{G}$ span characteristic values associated with distinct orbital configurations: purely radial orbits ($\beta_v \to 1$, $\mathcal{G} \to \infty$), isotropic orbits ($\beta_v = 0$, $\mathcal{G} = 3$), and purely tangential orbits ($\beta_v \to -\infty$, $\mathcal{G} \to 2$). The anisotropy parameter $\beta_v$ is related to the depletion factor $\mathcal{G}$ via $\beta_v = (\mathcal{G} - 3)/(\mathcal{G} - 2)$.

More recent studies \citep[e.g.,][]{Halenka2022,Rodriguez2024}, however, suggest that $\mathcal{G}$ may primarily reflect statistical undersampling of galaxy velocities near the caustic edges, rather than purely geometric effects from orbital anisotropy. Within the \clumiplus framework, we adopt an agnostic stance on the physical interpretation of $\mathcal{G}$, treating it as a free parameter to be empirically constrained from the data. Nevertheless, we use the original \citet{Diaferio1999} formulation to motivate a plausible prior range for $\mathcal{G}$ in our Bayesian analysis (Section~\ref{subsec:prior}).


\subsection{Focus on the Infall Region}

Our analysis focuses on the outer infall regions of galaxy clusters, specifically at $r > \Rinf \equiv 2 \Mpch$, a scale comparable to the $R_\mathrm{200\mathrm{m}}$ radius of high-mass clusters with $M_{200\mathrm{m}} \sim 10^{15}\Msunh$ \citep{Umetsu+Diemer2017}. In these regions, redshift diagrams provide robust sampling of tracer galaxies, enabling reliable measurements of the projected escape velocity \citep{Serra2011}. Cosmological simulations of hierarchical structure formation suggest that $\mathcal{G}$ varies only weakly with cluster-centric distance at these scales, remaining approximately constant owing to the coherent infall motion of surrounding galaxies \citep{Diaferio1999,Serra2011}. Crucially, as long as $\mathcal{G}$ is effectively constant at $r > \Rinf$, the \clumiplus algorithm yields robust inferences, independent of any specific interpretation of this depletion factor. 
Even when the velocity anisotropy parameter varies significantly with cluster radius, such as $\beta_v \in [0.4, 0.6]$ in cluster outskirts \citep[e.g.,][]{Wojtak2009,Serra2011,Biviano2013}, the corresponding variation in the caustic depletion factor remains modest, with $\mathcal{G} \in [3.7, 4.5]$.

Notably, the dynamical boundary of galaxy clusters---commonly referred to as the splashback radius, $\Rsp$---is predicted to be located at $\Rsp \sim R_{200\mathrm{m}}$ for accreting halos such as galaxy clusters \citep{Diemer+Kravtsov2014,More2015splash}. The splashback radius marks a physical transition separating the multistream intracluster region from the outer infall region \citep{1972ApJ...176....1G,Diemer+Kravtsov2014,More2015splash}. In these outer regions, the 3D density profile around halos follows a power-law form extending out to $r \sim 10\,R_{200\mathrm{m}}$ \citep{Diemer+Kravtsov2014,Diemer2023,Diemer2025}. This motivates our adoption of a projected power-law model to describe the external mass distribution beyond the lensing aperture (see Section~\ref{subsec:model}).

\section{Methodology: The CLUMI+ Algorithm}
\label{sec:clumiplus}

In this section, we introduce \clumiplus, a new methodology for modeling the mass distribution around galaxy clusters using a self-consistent, multiprobe framework. Building on the joint likelihood formalism developed in the original \clumi algorithm \citep{Umetsu+2011,Umetsu2013}, \clumiplus extends the approach to incorporate projected escape velocity constraints in addition to lensing observables. 

The underlying mass model characterizes the azimuthally averaged radial profile of the lensing convergence, $\kappa_\infty(\theta) = \Sigma(r_\perp)/\SigmaCritInf$, where $r_\perp = D_l \theta$ is the projected cluster-centric radius. Within the lensing field, $\kappa_\infty(\theta)$ is represented as a piecewise-constant function defined over concentric annuli, enabling a free-form reconstruction of the projected mass distribution. Beyond the lensing aperture, the model transitions to a projected power-law profile to describe the outer mass distribution. This choice is motivated by theoretical results from cosmological $N$-body simulations, which find that the 3D density profile of cluster halos follows an approximately power-law behavior in the infall region \citep{Diemer+Kravtsov2014,Diemer2023,Diemer2025}.

\subsection{Multiprobe Cluster Constraints}
\label{subsec:constraints}

The \clumiplus framework combines constraints from lensing and dynamical observations, extending the methodology of \textsc{clumi} \citep{Umetsu2013}. Specifically, \clumiplus jointly analyzes projected escape velocity measurements $\{\mathcal{A}_{i}\}_{i=1}^{\NEV}$ alongside weak-lensing observables: $\{\langle g_{+,i}\rangle\}_{i=1}^{\NWL}$ and $\{\langle n_{\mu,i}\rangle\}_{i=1}^{\NWL}$. Additionally, both \clumiplus and \clumi can incorporate strong-lensing aperture-mass constraints, $\{M_{\mathrm{ap},i}\}_{i=1}^{\NSL}$. 

Each weak-lensing profile is measured in $\NWL$ annular bins over the range $\theta \in [\thetamin^{(\mathrm{WL})}, \thetamax^{(\mathrm{WL})}]$ centered on the cluster. The total number of independent constraints is $N_\mathrm{data} = 2\NWL + \NSL + \NEV$, where the factor of 2 accounts for the inclusion of both shear and magnification constraints, $\NSL$ represents the number of available strong-lensing constraints, and $\NEV$ denotes the number of projected escape velocity constraints in the outer infall region at $r_\perp > \Rinf = 2 \Mpch$ (see Section~\ref{subsec:likelihood}).

The minimum radius for lensing constraints, $\thetamin$, is set by the availability of strong-lensing data. If strong-lensing constraints are available ($\NSL > 0$), we set $\thetamin = \thetamin^{(\mathrm{SL})}$. Otherwise, in the absence of strong-lensing constraints ($\NSL = 0$), the minimum radius is determined by weak-lensing data, $\thetamin = \thetamin^{(\mathrm{WL})}$.  The maximum radius for lensing constraints is $\thetamax=\thetamax^{(\mathrm{WL})}$.


\subsection{Mass Model}
\label{subsec:model}

The mass model in \clumiplus is described by a set of parameters, $\bm$, that characterize the azimuthally averaged projected mass distribution in terms of the lensing convergence $\kappa_\infty(\theta)$. This includes the average convergence within the innermost circular aperture of radius $\thetamin$, defined as
\begin{equation}
\overline{\kappa}_{\infty,\mathrm{min}} \equiv \overline{\kappa}_{\infty}(<\thetamin),
\end{equation}  
along with a piecewise-constant convergence profile, $\{\kappa_{\infty,i}\}_{i=1}^N$, defined over a set of $N$ annular bins covering the lensing aperture. For full details, see Appendix~\ref{appendix:model}.

Following the formalism of \citet{Umetsu2013}, we define the $(N+1)$-dimensional lensing signal vector as
\begin{equation}
\label{eq:bs}
\bs = 
  \left\{ 
  \overline{\kappa}_{\infty,\mathrm{min}},
  \kappa_{\infty,i}
  \right\}_{i=1}^N,
\end{equation}
where
\begin{equation}
N = \NSL + \NWL
\end{equation}
is the total number of radial bins spanning the interval $\theta \in [\thetamin, \thetamax]$, with $\NSL$ and $\NWL$ denoting the number of strong- and weak-lensing bins, respectively.

This piecewise-defined convergence model provides a flexible, free-form representation of the projected mass distribution within the lensing field. It enables direct predictions for the full set of lensing observables: the aperture masses $\{M_{\mathrm{ap},i}\}_{i=1}^{\NSL}$, the tangential reduced shear profile $\{\langle g_{+,i} \rangle\}_{i=1}^{\NWL}$, and the magnification bias profile $\{\langle n_{\mu,i} \rangle\}_{i=1}^{\NWL}$ (see Appendix~\ref{subsec:estimators} for details).

At this stage, the mass model adopted in \clumiplus is formally equivalent to that of the original \clumi algorithm \citep{Umetsu2013}, and serves as the foundation for incorporating additional dynamical constraints in the outer regions.

To extend the mass model beyond the radial range directly probed by lensing data, we include an external convergence component defined by a power-law profile:
\begin{equation}
\label{eq:kext}
\Theta(\theta - \thetamax)\,\kappa_{\infty,\mathrm{ext}}\left( \frac{\theta}{\thetamax} \right)^{-q},
\end{equation}
where $\kappa_{\infty,\mathrm{ext}}$ is the normalization at $\thetamax$, $q$ is the logarithmic slope, and $\Theta(x)$ is the Heaviside step function, defined as $\Theta(x) = 1$ for $x \ge 0$ and $\Theta(x) = 0$ otherwise. This term accounts for the contribution of mass beyond the lensing aperture and is essential for modeling the gravitational potential in the cluster infall regions, where phase-space information becomes increasingly informative.

In the \clumiplus framework, the complete set of model parameters, $\bm$, consists of the piecewise-defined convergence parameters $\bs$ describing the lensing field, the external power-law parameters $\kappa_{\infty,\mathrm{ext}}$ and $q$, and the depletion factor $\mathcal{G}$ associated with the escape velocity profile:
\begin{equation}
\bm = \left\{ \bs, \mathcal{G}, q, \kappa_{\infty,\mathrm{ext}} \right\}.
\end{equation}
This $(N + 4)$-dimensional parameter vector enables self-consistent predictions for both lensing and dynamical observables, bridging the cluster core and outer infall regions in a unified modeling framework.

\medskip

To evaluate the likelihood contribution from the projected escape velocity, we model the gravitational potential, $\Phi_\mathrm{N}(r)$, in the outer infall regions where caustic amplitude measurements provide additional constraints. As outlined below, we compute $\Phi_\mathrm{N}(r)$ from the underlying mass distribution under the assumption of spherical symmetry in the outer infall regions, allowing us to predict the escape velocity profile, $\vesc(r) = \sqrt{-2\Phi_\mathrm{N}(r)}$, for a given mass model $\bm$ (see Appendix~\ref{appendix:pred_vesc} for details).

The total mass enclosed within a 3D radius $r$ is related to the Newtonian potential via 
\begin{equation} 
\label{eq:Poisson} 
\Phi_\mathrm{N}(r) = -\int_r^{\infty}\! \frac{G M(<r')}{r'^2}\,dr' \simeq -\int_r^{r_\infty}\! \frac{G M(<r')}{r'^2} \, dr', 
\end{equation} 
where we truncate the integral at a finite outer boundary, $r_\infty = 20 \Mpch$, corresponding to approximately $10\,R_{200\mathrm{m}}$ for high-mass clusters. 
This scale is motivated by theoretical studies indicating a transition in modeling strategy \citep[see][]{Diemer+Kravtsov2014,Diemer2023}: beyond this radius, a simple power-law description of the infall region becomes inaccurate. At these large scales, the halo-model description based on the matter power spectrum (i.e., the 2-halo term) provides a more accurate representation of the density field. 
Nevertheless, for $r > 20 \Mpch$, the contribution to the cluster potential $\Phi_\mathrm{N}(r)$ from these distant structures is expected to be negligible and can be safely ignored.

To obtain $M(<r)$ from the projected mass distribution, we use the Abel integral transform \citep[][]{Binney2008,Umetsu+2011,Coles2014}: 
\begin{equation} 
\label{eq:abel} 
M(<r) = \pi r^2 \overline{\Sigma}(<r) - 4 \int_r^{r_\infty}\! R f\left( \frac{R}{r} \right) \Sigma(R) \, dR, 
\end{equation} 
where $f(x) = (x^2 - 1)^{-1/2} - \tan^{-1}(x^2 - 1)^{-1/2}$, $\overline{\Sigma}(<r)$ is the mean surface density within radius $r$, and $\Sigma(R)$ is the azimuthally averaged surface mass density.
The first term corresponds to the projected (cylindrical) mass within
radius $r$, $\Map(<r) = \pi r^2 \overline{\Sigma}(<r)$,
while the second term removes the contribution from mass elements at
$r'> r$, effectively deprojecting the cylindrical mass to obtain the
enclosed 3D mass.  In our numerical implementation, we truncate the integral at $r_\infty = 20 \Mpch$.

Notably, it is sufficient to evaluate the 3D mass profile $M(<r)$ at $r > \Rinf = 2 \Mpch$, where the escape velocity measurements $\{\mathcal{A}_i\}_{i=1}^{\NEV}$ are incorporated into the joint likelihood (Section~\ref{subsec:likelihood}). In contrast, the projected mass distribution $\Sigma(r_\perp)$ receives integrated contributions along the line of sight. Therefore, anchoring the outer 3D potential through velocity observables constrains the line-of-sight mass contributions and breaks degeneracies in the projected lensing observables. This coupling leads to improved stability and precision in the reconstruction of the convergence profile $\kappa_\infty(\theta)$, particularly in the outer halo regions.

\subsection{Calibration Parameters}
\label{subsec:calib}

\clumiplus accounts for calibration uncertainties in the multiprobe analysis through a set of nuisance parameters. These include the relative lensing efficiencies of the weak-lensing source populations, $\langle W\rangle_g$ and $\langle W\rangle_\mu$, along with the unlensed count normalization and logarithmic count slope for magnification bias, $\overline{n}_\mu$ and $\alpha$, respectively (Section~\ref{sec:caustic}):  
\begin{equation}
 \bc = \{\langle W\rangle_g, \langle W\rangle_\mu, \overline{n}_\mu, \alpha\}.
\end{equation}
These four parameters are treated as calibration nuisance parameters and marginalized over in the joint likelihood analysis. 

Combining the mass model and calibration nuisance parameters, the total number of free parameters in the joint likelihood analysis is $N+8$:  
\begin{equation}
\label{eq:fullmodel}
\bp = \left\{\bm, \bc\right\}.
\end{equation}

\subsection{Joint Likelihood Function}  
\label{subsec:likelihood}

In a Bayesian framework, we sample from the posterior probability density function of the model parameters, $\bp$, given the observed data, $\bd$, denoted as $P(\bp|\bd)$. The expectation values of any statistical function of $\bp$ converge to those derived from the marginalized posterior distribution, $P(\bp|\bd)$. Bayes's theorem relates the posterior probability to the prior and likelihood as  
\begin{equation}
 P(\bp|\bd) \propto P(\bp) P(\bd|\bp),
\end{equation}  
where $ P(\bp) $ is the prior distribution of the parameters and $ \mathcal{L}(\bp) \equiv P(\bd|\bp) $ is the likelihood function, describing the probability of the data given the model parameters.

For a joint analysis of weak-lensing and escape velocity measurements, the total likelihood function $ \mathcal{L}(\bp) $ is given by
\begin{equation} 
\mathcal{L}(\bp) = \mathcal{L}_\mathrm{WL}(\bp) \mathcal{L}_\mathrm{esc}(\bp),
\end{equation}  
where $\mathcal{L}_\mathrm{WL}(\bp)=\mathcal{L}_g(\bp)\mathcal{L}_\mu(\bp)$.
Here $\mathcal{L}_g(\bp)$, $\mathcal{L}_\mu(\bp)$, and $\mathcal{L}_\mathrm{esc}(\bp)$ correspond to the likelihood functions for the observables $\{\langle g_{+,i}\rangle\}_{i=1}^{\NWL}$, $\{\langle n_{\mu,i}\rangle\}_{i=1}^{\NWL}$, and $\{\mathcal{A}_{i}\}_{i=1}^{\NEV}$, respectively.

If strong-lensing aperture-mass constraints $\{M_{\mathrm{ap},i}\}_{i=1}^{\NSL}$ are available, the total likelihood function is extended to  
\begin{equation} 
\mathcal{L}(\bp) = \mathcal{L}_\mathrm{WL}(\bp)  
 \mathcal{L}_\mathrm{SL}(\bp) 
 \mathcal{L}_\mathrm{esc}(\bp).
\end{equation}  
where $\mathcal{L}_\mathrm{SL}(\bp)$ represents the contribution from strong-lensing constraints.

The corresponding log-likelihoods, up to constant terms, are given by  
\begin{eqnarray}
\ln\mathcal{L}_{g}(\bp) &=& -\frac{1}{2}\sum_{i=1}^{\NWL}\frac{[\langle g_{+,i}\rangle-\widehat{g}_{+,i}(\bp)]^2}{\sigma^2_{+,i}},\\
\ln\mathcal{L}_\mu(\bp) &=& -\frac{1}{2}\sum_{i=1}^{\NWL}\frac{[\langle n_{\mu,i}\rangle-\widehat{n}_{\mu,i}(\bp)]^2}{\sigma^2_{\mu,i}},\\
\label{eq:Lsl}
\ln\mathcal{L}_\mathrm{SL}(\bp)&=&-\frac{1}{2} \sum_{i=1}^{\NSL}\frac{[M_{\mathrm{ap},i}-\widehat{M}_{\mathrm{ap},i}(\bp)]^2}{\sigma^2_{\mathrm{ap},i}},\\
\label{eq:Lev}
\ln\mathcal{L}_\mathrm{esc}(\bp) &=&-\frac{1}{2} \sum_{i=1}^{\NEV}\frac{[\mathcal{A}_{i}-\widehat{\mathcal{A}}_i(\bp)]^2}{\sigma^2_{\mathcal{A},i}},
\end{eqnarray}  
where $\widehat{g}_{+}$, $\widehat{n}_\mu$, $\widehat{M}_\mathrm{ap}$, and $\widehat{\mathcal{A}}$ are the theoretical predictions for the respective observables, each determined by the model parameters $\bp = \{\bm,\bc\}$. A detailed description of these estimators is provided in Appendix~\ref{subsec:estimators}.

The uncertainty $\sigma_{+,i}$ for $\langle g_{+,i}\rangle$ primarily arises from intrinsic ellipticity variance and shape measurement errors \citep{Umetsu2020rev}. The uncertainty $\sigma_{\mu,i}$ for $\langle n_{\mu,i}\rangle$ accounts for both Poisson counting errors and variance due to the intrinsic clustering of the source population \citep{Umetsu2022}. The strong-lensing aperture-mass estimates $M_{\mathrm{ap},i}$ and their uncertainties $\sigma_{\mathrm{ap},i}$ are derived from detailed modeling of the deflection field, $\bnabla\psi(\btheta)$, based on sets of multiply lensed images \citep{Umetsu2016clash}.  

Additionally, we extract measurements of the projected escape velocity, $\{\mathcal{A}_i\}_{i=1}^{\NEV}$, from the caustic amplitude profile $\mathcal{A}(r_\perp)$ at $r_\perp > \Rinf = 2 \Mpch$, as identified in redshift diagrams using the caustic technique. To minimize covariance between radial bins, we sample data points $\mathcal{A}_i$ at uniform intervals along the $r_\perp$ axis, setting the interval to match the median radial smoothing length $\overline{h}_r$ of the adaptive kernel used to construct the continuous phase-space density distribution \citep{Pizzardo2023ct}. Crucially, we account for a 20\percent scatter due to projection effects by incorporating it in quadrature into the uncertainty $\sigma_{\mathcal{A},i}$ as this contribution often dominates the total error budget \citep{Serra2011}.

Finally, we note that the \clumiplus framework, like \clumi, is readily extensible to ensemble cluster analyses, enabling joint modeling of stacked weak-lensing and phase-space profiles for statistical samples of galaxy clusters.

\subsection{Priors}
\label{subsec:prior}

\clumiplus adopts uninformative, uniform priors for all model parameters. This choice ensures that the inference is driven primarily by the data, with minimal reliance on prior assumptions. As a result, the model remains flexible and robust, provided that the input data vectors used in the joint analysis are themselves unbiased.

The uniform priors for the mass model parameters $\bm$ are specified as follows:
\begin{equation}
 \begin{aligned}
  \overline{\kappa}_\infty(<\theta_\mathrm{min}) &\in [0,5],\\ 
  \kappa_\infty(\theta_i) &\in 
    \begin{cases} 
      [0,5], & \text{for } \theta_i \in [\thetamin^{(\mathrm{SL})},\thetamax^{(\mathrm{SL})}],\\
      [0,1], & \text{for } \theta_i \in [\thetamin^{(\mathrm{WL})},\thetamax^{(\mathrm{WL})}].
    \end{cases} \\
  \kappa_{\infty,\mathrm{ext}} &\in [0,1],\\
  q &\in [0,2],\\
  {\cal G} &\in [2, 15],
 \end{aligned}
\end{equation}
where \clumiplus assumes that weak-lensing measurements $\{\langle g_{+,i}\rangle, \langle n_{\mu,i}\rangle\}_{i=1}^{\NWL}$ are obtained in subcritical regions with $\kappa_{\infty}(\theta) < 1$ (Section~\ref{sec:lens}). The upper prior bound of $5$ on the central convergence parameter and the piecewise-defined convergence parameters in the strong-lensing regime is chosen to conservatively encompass the range expected for cluster-scale lenses \citep[e.g.,][]{Merten2015clash}.

A uniform prior on the depletion factor $\mathcal{G}$ is mathematically natural and optimal for unbiased inference of the gravitational potential, since $-2\Phi_\mathrm{N} = \mathcal{G} \mathcal{A}^2$ under the assumption that $\mathcal{A}(r_\perp)$ is accurately measured. The prior range adopted for $\mathcal{G}$ spans a factor of $7.5$ in gravitational potential magnitude at fixed $\mathcal{A}$, thereby allowing for a broad range of possible line-of-sight mass contributions. 

In the geometric interpretation of $\mathcal{G}$ as a correction for orbital velocity anisotropy (Section~\ref{sec:caustic}), this prior maps nonlinearly to a range in the velocity anisotropy parameter, $\beta_v$ \citep{Wojtak2009}, extending from $-\infty$ to $\approx 0.92$. This range broadly encompasses the spectrum of orbital configurations expected for cluster-scale halos \citep[][]{Serra2011,Lemze2012,He2024}, ensuring that the prior on $\mathcal{G}$ remains sufficiently unrestrictive for empirical inference.

Additionally, we marginalize over the calibration parameters $\bc$ using uniform priors, thereby propagating uncertainties associated with the observational characterization of source galaxy populations in shear and magnification measurements (Section~\ref{subsec:calib}).

\subsection{Implementation}
\label{subsec:implementation}

We implement \clumiplus using a Markov Chain Monte Carlo (MCMC) approach with Metropolis--Hastings sampling and convergence diagnostics, following \citet{Umetsu+2011}. As summary statistics for each model parameter, we adopt the biweight estimator of \citet{1990AJ....100...32B} to characterize the central location of marginalized one-dimensional (1D) posterior distributions \citep[][]{Sereno+Umetsu2011}. 

Biweight statistics offer robustness against outliers by assigning higher weights to data points closer to the central location of the distribution, thereby reducing the influence of extreme values \citep{1990AJ....100...32B}. The biweight center provides a reliable measure of central tendency, particularly for skewed or non-Gaussian distributions. Unlike the maximum a posteriori estimator, which corresponds to the mode, the biweight center approximates the median in the presence of asymmetry, as seen in lognormal-like posteriors typical of low signal-to-noise ($\mathrm{S/N}$) lensing bins (Section~\ref{subsec:tng_results}).

The statistical covariance matrix $C^\mathrm{stat}$ for the piecewise-defined convergence parameters (Equation~\ref{eq:bs}) is estimated from the posterior MCMC samples as
\begin{equation}
\label{eq:cstat} 
\left(C^\mathrm{stat}\right)_{ij} = \mathrm{Cov}(s_i, s_j) = \langle (s_i - \langle s_i \rangle)(s_j - \langle s_j \rangle) \rangle.
\end{equation}

%

\section{Validating CLUMI+ with Synthetic Observations}
\label{sec:tng}

We validate \clumiplus using synthetic observations of a massive galaxy cluster extracted from the IllustrisTNG suite of cosmological hydrodynamical simulations \citep{Pillepich18,Springel18,Nelson19}. The goal is to assess the robustness of the \clumiplus framework under realistic cluster conditions, including triaxiality, substructure, and projection effects.

Specifically, the validation tests examine whether the algorithm can recover unbiased estimates of the projected surface mass density $\Sigma(r_\perp)$ and the 3D enclosed mass profile $M(<r)$ over the radial domain probed by multiprobe data, using synthetic weak-lensing and phase-space observables derived from the simulation. In \clumiplus, escape velocity information is incorporated under three working assumptions (Sections~\ref{sec:caustic} and \ref{sec:clumiplus}):
\begin{enumerate}
\item a Newtonian interpretation of the escape velocity;
\item constancy of the caustic depletion factor, $\mathcal{G}$, in the infall region ($r > \Rinf$); and
\item spherical symmetry of the 3D mass distribution in the infall region ($r > \Rinf$),
\end{enumerate}
where we adopt $\Rinf = 2 \Mpch$ throughout this study.

Although these assumptions are not tested independently, accurate recovery of both projected and 3D mass profiles provides empirical validation of the \clumiplus framework under realistic conditions.

\begin{figure}[htb!] 
 \begin{center}
  \includegraphics[width=0.9\columnwidth,angle=0,clip]{\FIG/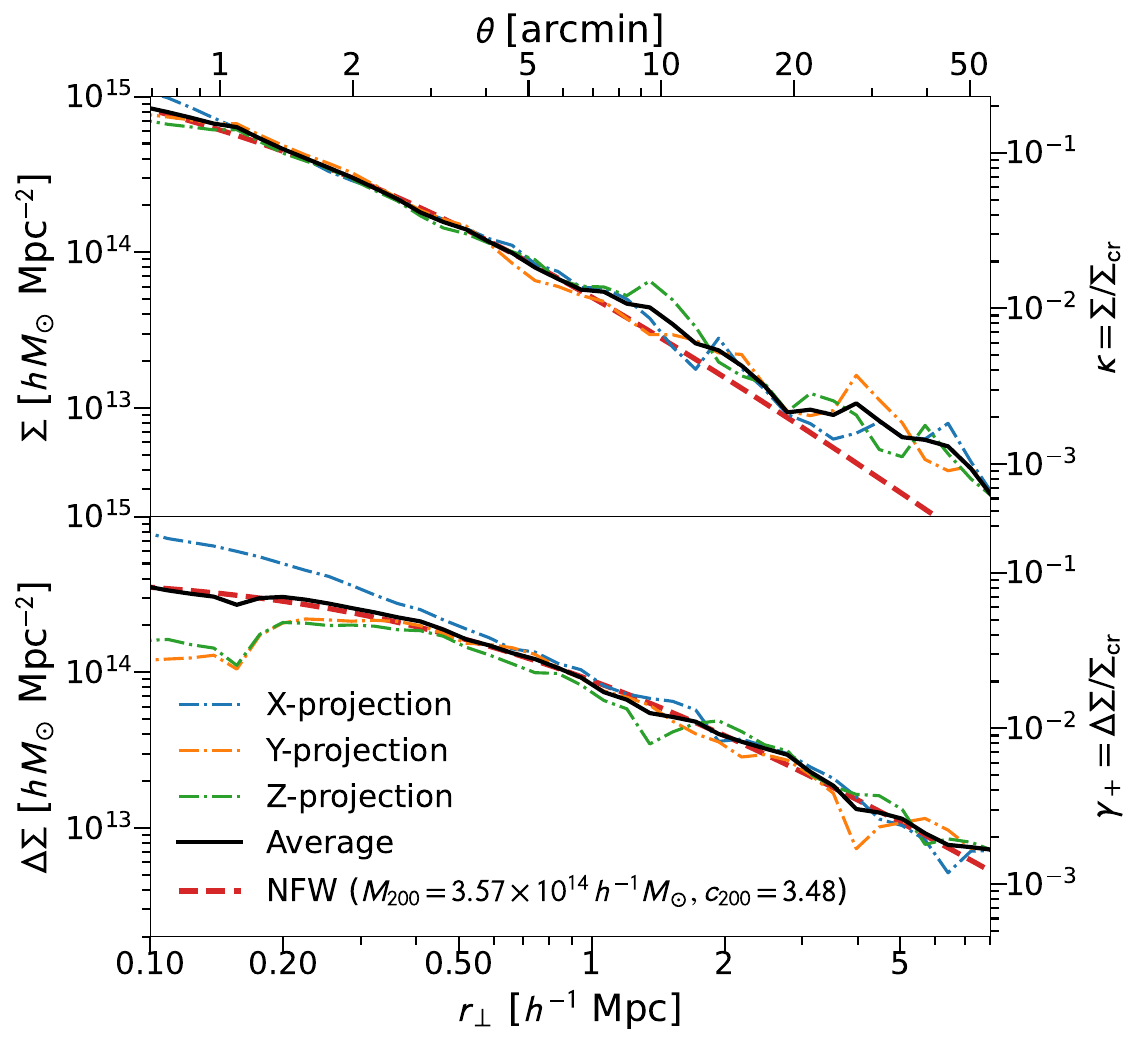}
 \end{center}
\caption{Azimuthally averaged lensing profiles (dotted-dashed lines) of a massive TNG300-1 cluster at $\zl=0.21$ centered on the BCG projected along the $X$ (blue), $Y$ (orange), and $Z$ (green) directions. These profiles are constructed from the projected mass maps shown in Appendix~\ref{appendix:tng} (Figure~\ref{fig:tng_maps}). The top and bottom panels show the surface mass density, $\Sigma(r_\perp)=\SigmaCrit\, \kappa(\theta)$, and the excess surface mass density, $\Delta\Sigma(r_\perp)=\SigmaCrit\, \gamma_+(\theta)$, respectively. In each panel, the black solid line represents the average profile over all three projections. The red dashed line in each panel shows the projected NFW profile corresponding to the cluster mass and concentration.
\label{fig:tng_lens}}
\end{figure}

\subsection{IllustrisTNG Simulations}
\label{subsec:tng-info}

The IllustrisTNG suite of cosmological hydrodynamical simulations \citep{Pillepich18,Springel18,Nelson19} is a set of gravo-magnetohydrodynamical simulations based on the $\Lambda$CDM model. The simulations are initialized at $z = 127$ using \Planck cosmological parameters \citep{ade2016planck}: cosmological constant density $\OL = 0.6911$, matter density $\Om = 0.3089$, baryon density $\Ob = 0.0486$, Hubble constant $H_0 = 67.74$~km~s$^{-1}$~Mpc$^{-1}$, power spectrum normalization $\sigma_8 = 0.8159$, and scalar spectral index $n_s = 0.9667$.

All IllustrisTNG baryonic simulations are performed using the AREPO code \citep{Springel10}, which solves the equations of continuum magnetohydrodynamics coupled with Newtonian self-gravity. The simulations track the evolution of four matter components: dark matter, gas, stars, and black holes.

We use the TNG300-1 run of IllustrisTNG, the highest-resolution baryonic simulation among those with the largest simulated volumes. The comoving box size is $302.6$~Mpc per side. TNG300-1 contains $2500^3$ dark matter particles with mass $m_\mathrm{DM} = 5.88 \times 10^{7} M_{\odot}$ and an equal number of gas cells with an average mass of $m_\mathrm{b} = 1.10 \times 10^{7} M_{\odot}$.  

IllustrisTNG provides high-fidelity synthetic observations in a controlled, realistic environment, making it well-suited for testing the key assumptions underlying the \clumiplus algorithm. 
For this analysis, we selected a massive halo from TNG300-1, specifically the sixth most massive halo at a redshift of $z_\mathrm{cl} = 0.21$ (FOF ID \#3). This halo has a mass of $M_{200\mathrm{c}} = 3.57 \times 10^{14} \Msunh$ and a concentration of $c_{200\mathrm{c}} = 3.48$ \citep{Pizzardo2023ct}. It exhibits a complex configuration in its outer infall region, including two additional infalling halos: one at cluster scale and one at group scale. The combination of a well-characterized mass distribution and realistic projection effects from infalling substructures makes this system an ideal test bed for evaluating the performance and robustness of \clumiplus under controlled conditions.

Analyses were performed along three orthogonal projections ($X$-, $Y$-, and $Z$-axes), each centered on the brightest cluster galaxy (BCG), identified as the most massive subhalo within the halo. Caustic analyses were then applied to synthetic redshift diagrams generated from synthetic galaxy catalogs, where ``galaxies'' are defined as Subfind subhalos with stellar masses above $10^8 M_\odot$.

\subsection{Synthetic Weak-lensing Observations}
\label{subsec:tng_wl}

\subsubsection{Projected Mass Maps}
\label{subsubsec:massmaps}

We generate a 2D projected mass map for each of the three orthogonal projections of the TNG300-1 cluster. To construct each synthetic mass map, we project the 3D positions of the matter elements along the corresponding projection axis and sum their masses in 2D bins of celestial coordinates. Below, we outline our procedure in detail.

We define a cubic volume centered on the cluster, with a side length of 20~Mpc ($\simeq 24.2$~comoving Mpc) at $z_\mathrm{cl}=0.21$. From the raw simulation data, we extract the 3D coordinates of all matter elements within this volume, including all four matter species implemented in TNG300-1. Each set of 3D coordinates is then projected along a selected axis. We define a reference frame in which the observer is positioned along the projection axis, with the cluster center as the origin. The original comoving 3D coordinates of the matter elements, $r_{\mathrm{cl},i}$, are measured relative to the cluster center.

The comoving distance between the cluster center and the observer is given by
\begin{equation} 
r_\mathrm{cl} = \frac{c}{H_0} \int_{0}^{z_\mathrm{cl}}\! \frac{dz}{E(z)},
\end{equation} 
where $E(z) = \left[\Om(1+z)^3 + \OL \right]^{1/2}$ is the dimensionless expansion rate in a flat $\Lambda$CDM cosmology.

The transformed 3D positions are then given by  
\begin{equation} 
\br_i = \br_\mathrm{cl} + \br_{\mathrm{cl},i},
\end{equation} 
where $\br_\mathrm{cl}$ is directed along the projection axis and connects the observer to the cluster center. Setting the celestial coordinates of the cluster center to $(\alpha_\mathrm{cl}, \delta_\mathrm{cl}) = (\pi/2,0)$, standard geometrical transformations yield the celestial coordinates $(\alpha_i,\delta_i)$ of the matter elements.  
We emphasize that this projection was performed in spherical celestial coordinates without assuming the flat-sky approximation. In contrast, the \clumiplus modeling assumes a flat-sky geometry, which is accurate for the angular scales relevant to cluster weak-lensing analyses and those adopted in this work.

\begin{figure}[htb!] 
 \begin{center}
  \includegraphics[width=0.9\columnwidth,angle=0,clip]{\FIG/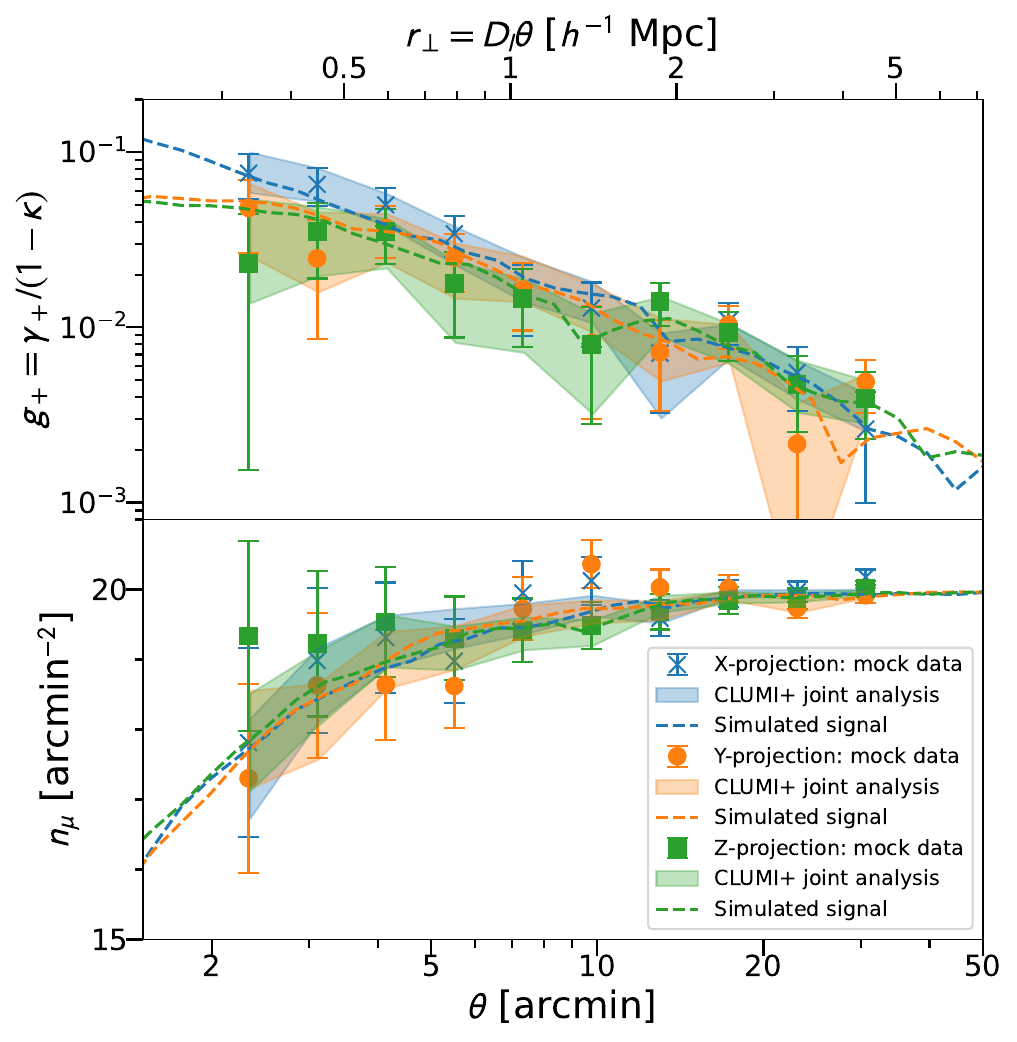}\\
 \end{center}
\caption{Synthetic weak-lensing observations of the TNG300-1 cluster constructed along the $X$ (blue), $Y$ (orange), and $Z$ (green) projections, centered on the BCG. The top and bottom panels show the reduced tangential shear, $g_+(\theta)$, and the magnified source counts, $n_\mu(\theta)$, respectively. Symbols with error bars show the synthetic measurements in each bin. Colored dashed lines show the corresponding noise-free input lensing profiles for each projection (Figure~\ref{fig:tng_lens}). For each observed profile, the shaded regions represent the marginalized $1\sigma$ confidence intervals from the joint \clumiplus analysis of synthetic weak-lensing and phase-space data (Figure~\ref{fig:tng_mock_vesc}).
\label{fig:tng_mock_wl}
}
\end{figure}

We define a region in celestial coordinates centered at $(\alpha_\mathrm{cl}, \delta_\mathrm{cl})$ with a total angular extent of  
\begin{equation}
2\Delta \theta = 2\arcsin \left[ \frac{(1+z_\mathrm{cl})R_\mathrm{sim}}{r_\mathrm{cl}} \right]
\end{equation}  
along each angular dimension, where $R_\mathrm{sim}=10\Mpc \approx 6.7\Mpch$.
This region corresponds to a projected square area of 10~Mpc per side at $\zl = z_\mathrm{cl} = 0.21$. We divide this region into square 2D angular bins of size $\delta\theta = 5\times 10^{-5}$~rad $\approx 10.3\arcsec$, yielding $N_\theta \equiv 2\Delta\theta/\delta\theta +1 = 541$ bins per dimension and a total of $N_\theta^2 = 292,681$ 2D angular bins. 

Finally, we assign each matter element to its corresponding 2D angular bin based on its celestial coordinates and sum the masses of all elements within each bin. This procedure produces one projected mass map for each of the three projections under consideration. The resulting mass maps are presented in Appendix~\ref{appendix:tng} (see Figure~\ref{fig:tng_maps}).  

Figure~\ref{fig:tng_lens} shows the azimuthally averaged surface mass density, $\Sigma(r_\perp)$, and excess surface mass density, $\Delta\Sigma(r_\perp)$, of the TNG300-1 cluster centered on the BCG, projected along three orthogonal directions. The average profile over all three projections is shown as a black solid line, which closely follows the projected Navarro--Frenk--White (\citealt{NFW1996}; hereafter NFW) profile corresponding to the cluster mass and concentration, $M_{200\mathrm{c}} = 3.57 \times 10^{14} \Msunh$ and $c_{200\mathrm{c}} = 3.48$ at $\zl = 0.21$. This halo mass corresponds to an overdensity radius of $R_{200\mathrm{c}}=1.07\Mpch$.  At $r_\perp \simgt 3\Mpch$, $\Sigma(r_\perp)$ systematically exceeds the NFW prediction, while $\Delta\Sigma(r_\perp)$ remains consistent with it. This behavior arises because the clustering contribution, well approximated by a constant surface mass density over $r_\perp \simlt 5 R_{200\mathrm{c}}$ \citep{Oguri+Hamana2011}, becomes significant at these large scales.

\subsubsection{Cluster Lensing Profiles}
\label{subsubsec:tng_mock_profiles}

For each projection of the TNG300-1 cluster, we create synthetic weak-lensing data, $\{\langle g_{+,i}\rangle,  \langle n_{\mu,i}\rangle\}_{i=1}^{\NWL}$, using $\NWL=10$ logarithmically spaced bins over the range $\theta \in [\thetamin,\thetamax] = [2\arcmin, 35\arcmin]$. The maximum lensing radius corresponds to $D_l \thetamax \approx 5\Mpch$, which lies within the projected simulation extent $R_\mathrm{sim}$ (Section~\ref{subsubsec:massmaps}). This configuration enables a direct assessment of the recovery of the external convergence term from the synthetic data.

Both source populations used for shear and magnification measurements are placed at $\zs=1.2$, with an unlensed number density of $n_s = 20$~arcmin$^{-2}$, comparable to the depth of deep ground-based observations \citep[Figure~\ref{fig:a2261_wl} in Appendix~\ref{appendix:a2261_data}; see also][]{Umetsu2014clash,Umetsu2015}. Synthetic shear and magnification profiles are generated by adding random Gaussian noise to each radial bin of the respective true input profile (Equations~(\ref{eq:gt}) and (\ref{eq:magbias})) derived from the TNG300-1 simulation (Figure~\ref{fig:tng_lens}). We assume that sources are randomly distributed, neglecting intrinsic source clustering. Reduced tangential shear data include random shape noise with an intrinsic dispersion of $\sigma_g=0.3$ per shear component. For magnification bias, we assume a maximally depleted population ($\alpha=0$) and model statistical uncertainties in source counts using a Gaussian approximation with a standard deviation of $\sqrt{N_s}$, where $N_s$ is the expected number of sources per radial bin. Since $N_s \gg 10$ in our setup, this approximation is accurate for modeling Poisson fluctuations. We note that these optimistic modeling assumptions, namely the use of a maximally depleted source population ($\alpha=0$) and unclustered sources, were adopted to ensure sufficient $\mathrm{S/N}$s for assessing projection effects on lensing observables. These choices allow us to evaluate the response of lensing profiles in a controlled setting, particularly their sensitivity to halo triaxiality at small cluster radii \citep[e.g.,][]{Umetsu2020rev}.

To isolate the performance of the \clumiplus algorithm and assess its underlying assumptions, we assume perfect knowledge of the calibration parameters, $\bc = \{\langle W\rangle_g, \langle W\rangle_\mu, \overline{n}_\mu, \alpha\}$, with $\overline{n}_\mu = n_s$, and set $f_g = 1$.

Figure~\ref{fig:tng_mock_wl} presents the synthetic weak-lensing measurements, $\{\langle g_{+,i}\rangle\}_{i=1}^{\NWL}$ and $\{\langle n_{\mu,i}\rangle\}_{i=1}^{\NWL}$, for each of the three orthogonal projections of the TNG300-1 cluster. These are shown alongside the corresponding noise-free input lensing profiles. For the tangential distortion signal, the total detection $\mathrm{S/N}$s are $\mathrm{S/N} = 9.9$, $7.7$, and $7.7$ for the $X$-, $Y$-, and $Z$-projections, respectively. For the magnification bias signal, the corresponding $\mathrm{S/N}$ values are $3.6$, $4.4$, and $3.1$, respectively. The combined $\mathrm{S/N}$ values for the weak-lensing profiles are therefore of the order of $\sim 10$.

\subsection{Synthetic Phase-Space Observations}
\label{subsec:tng_caustic}

\subsubsection{Redshift Diagrams}
\label{subsubsec:tng_vr}

We construct redshift diagrams for each projection of the TNG300-1 cluster, representing the 2D projected phase space of tracer galaxies around the cluster. These diagrams plot the line-of-sight velocity of galaxies relative to the cluster center, $v_\mathrm{los}$, against the projected cluster-centric radius, $r_\perp$. The resulting $v_\mathrm{los}$--$r_\perp$ diagram serves as the foundation of the caustic technique used to identify caustic curves, which trace the projected escape velocity profile, $\mathcal{A}(r_\perp)$.  

Following \citet[][see their Section~3.3]{Pizzardo2023ct}, we generate these diagrams using synthetic catalogs of simulated cluster galaxies, which provide R.A., decl., and total redshift along the line of sight. To ensure consistency with our mass map analysis (Section~\ref{subsubsec:massmaps}), we include all simulated galaxies within the same cubic volume used for the projected mass distribution.  

We apply the caustic technique to determine $\mathcal{A}(r_\perp)$ for each of the three synthetic $v_\mathrm{los}$--$r_\perp$ diagrams. The results are presented in Appendix~\ref{appendix:tng} (see Figure~\ref{fig:tng_caustic}).

\begin{figure}[htb!] 
 \begin{center}
  \includegraphics[width=0.9\columnwidth,angle=0,clip]{\FIG/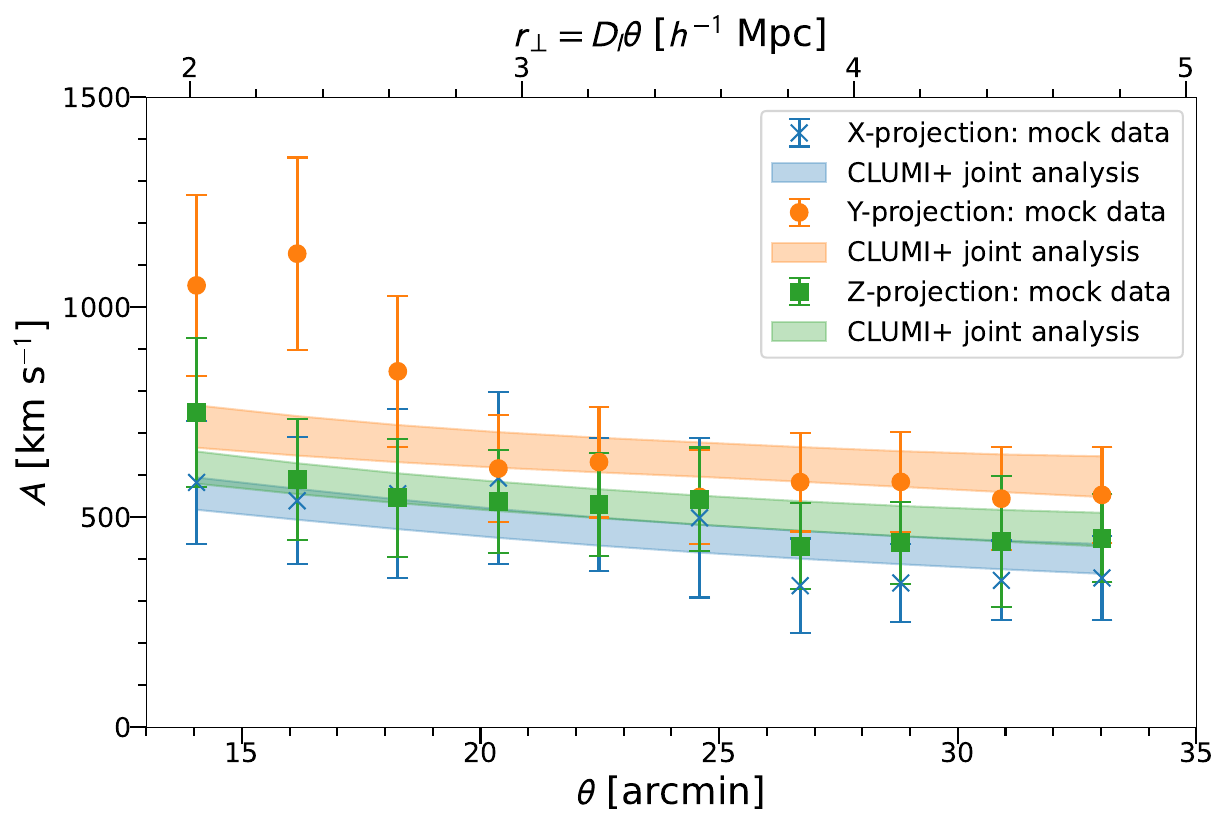}
 \end{center}
\caption{Same as Figure~\ref{fig:tng_mock_wl}, but for synthetic phase-space observations of the TNG300-1 cluster created along the $X$ (blue), $Y$ (orange), and $Z$ (green) directions, centered on the BCG. Symbols with error bars represent synthetic measurements of the projected escape velocity $\mathcal{A}(r_\perp)$ in each radial bin obtained using the caustic technique (see Figure~\ref{fig:tng_caustic}). For each observed profile, the colored shaded region denotes the marginalized $1\sigma$ confidence interval from the joint \clumiplus analysis of synthetic weak-lensing and phase-space measurements (see also Figure~\ref{fig:tng_mock_vesc}).  
\label{fig:tng_mock_vesc}}
\end{figure}

\begin{figure*}[htb!] 
 \centering
 \includegraphics[width=0.9\columnwidth,angle=0,clip]{\FIG/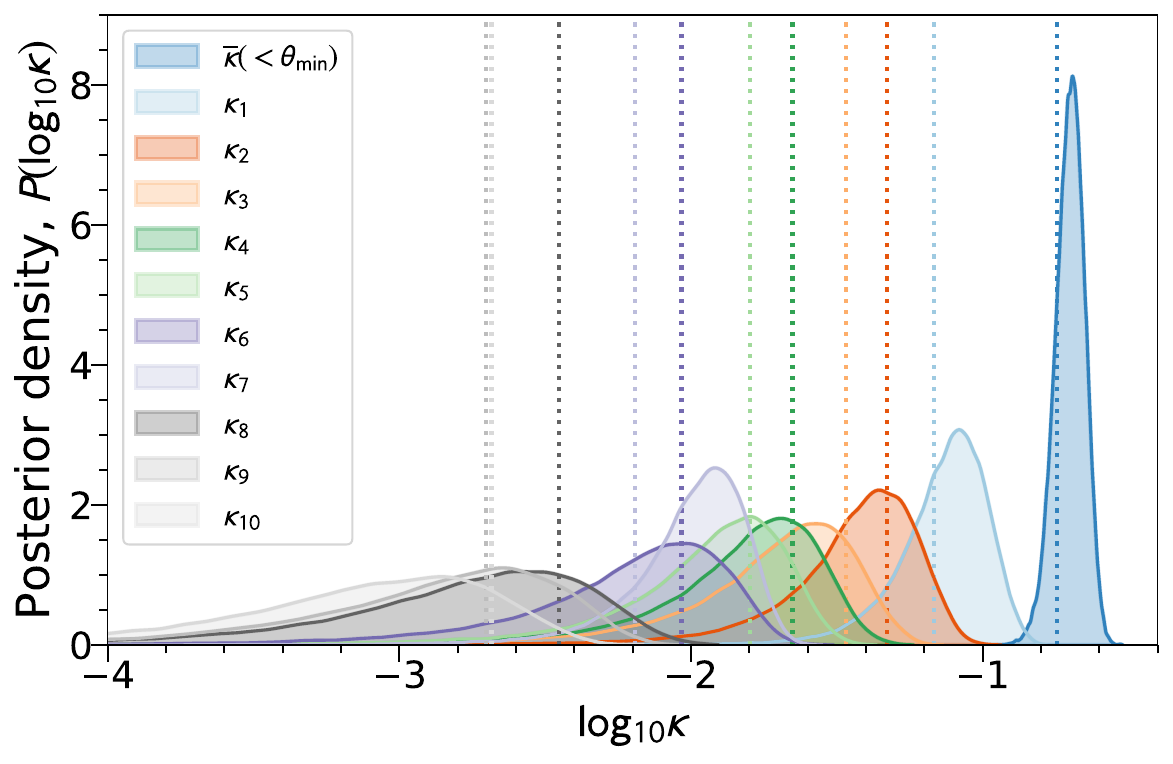}
 \includegraphics[width=1.1\columnwidth,angle=0,clip]{\FIG/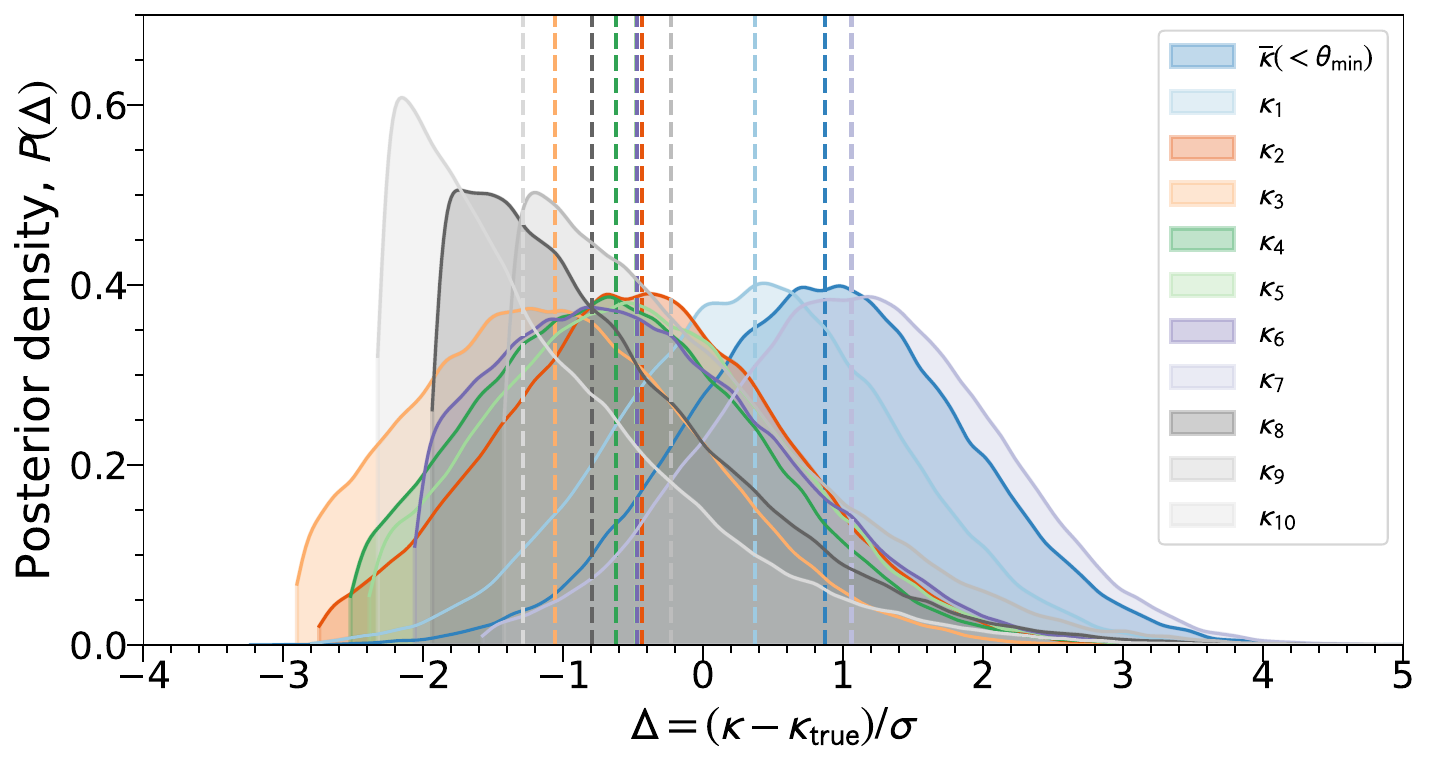}
\caption{Marginalized posterior probability distributions of the piecewise-constant convergence parameters, $\bs = \left\{ \overline{\kappa}_{\infty,\mathrm{min}}, \kappa_{\infty,i} \right\}_{i=1}^{N}$ with $N = 10$, inferred using \clumiplus from synthetic observations (Figures~\ref{fig:tng_mock_wl} and \ref{fig:tng_mock_vesc}). Results are shown for the $X$-projection. The left panel displays the distributions of the log-transformed parameters, $P(\log_{10} \kappa_\infty)$, with each parameter color-coded and shaded. Dotted vertical lines mark the true values. The right panel shows the same distributions as fractional deviations, $\Delta = (\kappa_\infty - \kappa_{\infty,\mathrm{true}})/\sigma$, normalized by the standard deviation $\sigma$ of each distribution. Dashed vertical lines indicate the biweight center of each posterior.
\label{fig:tng_kappa_pdf}}
\end{figure*}

\subsubsection{Projected Escape Velocity Profiles}
\label{subsubsec:tng_mock_vesc}

The caustic technique applies a Gaussian kernel to transform the discrete 2D projected phase space of galaxies into a continuous distribution \citep{Diaferio1999,Serra2011}. The optimal local smoothing length, $h_r$, is determined by minimizing the difference between the continuous and discrete distributions. We find median values of $\overline{h}_r = 230\kpch$, $120\kpch$, and $140\kpch$ for the $X$-, $Y$-, and $Z$-projections, respectively. To ensure consistency in our \clumiplus analyses across all projections, we adopt the largest value, $\overline{h}_r = 230\kpch$, as the sampling interval for extracting data points along the $r_\perp$ axis (Section~\ref{subsec:likelihood}). We extract caustic amplitudes over the range $r_\perp\in [2, 5]\Mpch$ at a uniform interval of $230\kpch$, yielding $\NEV=10$ data points per projection, $\{\mathcal{A}_i\}_{i=1}^{\NEV}$.

Figure~\ref{fig:tng_mock_vesc} shows the resulting $\mathcal{A}(r_\perp)$ profiles. The total detection $\mathrm{S/N}$ values are $10.6$, $15.3$, and $13.1$ for the $X$-, $Y$-, and $Z$-projections, respectively. 
At $r_\perp > 2\Mpch$, where projected escape velocity information is incorporated into \clumiplus, the profiles from the $X$- and $Z$-projections agree within $\sim 10\percent$ on average. However, the $Y$-projection exhibits systematically higher caustic amplitudes, exceeding the other two projections by $\sim 40\percent$ on average, which is about twice the 20\percent scatter level found in cosmological simulations \citep{Serra2011} and adopted in this work.

This discrepancy is attributed to two infalling substructures along the $Y$-projection, positioned on opposite sides of the cluster center at 3D cluster-centric distances greater than $5 R_{200\mathrm{c}}$. One substructure is approaching the observer while the other is receding, and their internal velocity dispersions contribute to the broadening of the $v_\mathrm{los}$--$r_\perp$ distribution (see Figure~\ref{fig:tng_caustic} in Appendix~\ref{appendix:tng}). This artificial thickening systematically enhances the estimated caustic amplitude, particularly around $r_\perp \sim 2.3\Mpch$ (Figure~\ref{fig:tng_mock_vesc}).

\begin{figure}[htb!] 
 \begin{center}
  \includegraphics[width=0.9\columnwidth,angle=0,clip]{\FIG/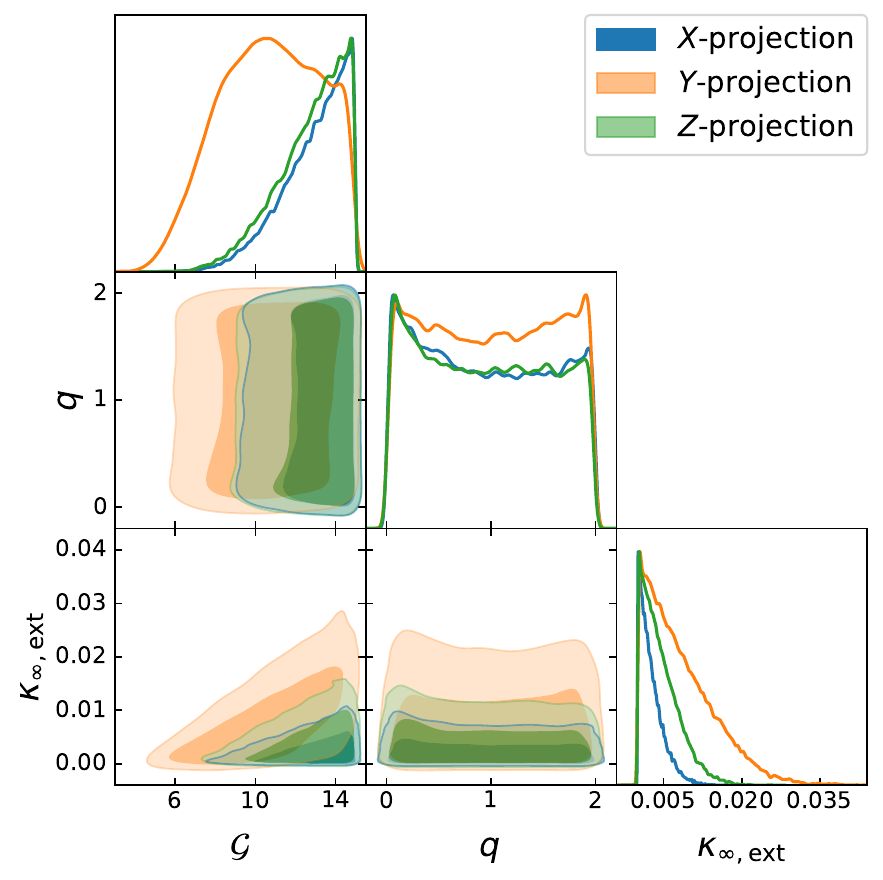}
 \end{center}
\caption{Marginalized posterior distributions (68\percent and 95\percent confidence levels) for a subset of the model parameters, $(\mathcal{G}, q, \kappa_{\infty,\mathrm{ext}})$, inferred using \clumiplus from synthetic observations (Figures~\ref{fig:tng_mock_wl} and \ref{fig:tng_mock_vesc}). Results are shown for the $X$ (blue), $Y$ (orange), and $Z$ (green) projections. Diagonal panels show the 1D marginalized posterior distributions for each parameter, while the off-diagonal panels display the corresponding 2D joint posterior contours.
\label{fig:tng_triangle}}
\end{figure}

\subsection{Validation Results}
\label{subsec:tng_results}

Following the procedures outlined in the preceding subsections, we apply \clumiplus to perform a joint likelihood analysis of synthetic observations of the TNG300-1 cluster. The analysis incorporates the synthetic weak-lensing profiles $\{\langle g_{+,i}\rangle, \langle n_{\mu,i}\rangle\}_{i=1}^{\NWL}$ (Figure~\ref{fig:tng_mock_wl}) and projected escape velocity measurements $\{\mathcal{A}_i\}_{i=1}^{\NEV}$ at $r_\perp > \Rinf = 2\Mpch$ (Figure~\ref{fig:tng_mock_vesc}), using the methodology detailed in Section~\ref{sec:clumiplus}. 

For this validation, we adopt $\NWL = \NEV = N = 10$ bins for each profile. The mass model $\bm$ in \clumiplus is therefore defined by $N + 4 = 14$ parameters, including the piecewise-constant convergence profile and a power-law extension that models the external convergence beyond the lensing aperture. The total $\mathrm{S/N}$ values of the weak-lensing and escape velocity data are $\sim10$ and $\sim13$, respectively (Sections~\ref{subsubsec:tng_mock_profiles} and \ref{subsubsec:tng_mock_vesc}). This approximate balance in constraining power indicates that the joint modeling of lensing and dynamical information is statistically well matched and jointly informative.

\begin{figure*}[htb!] 
  \begin{center}
   \includegraphics[width=1.04\columnwidth,angle=0,clip]{\FIG/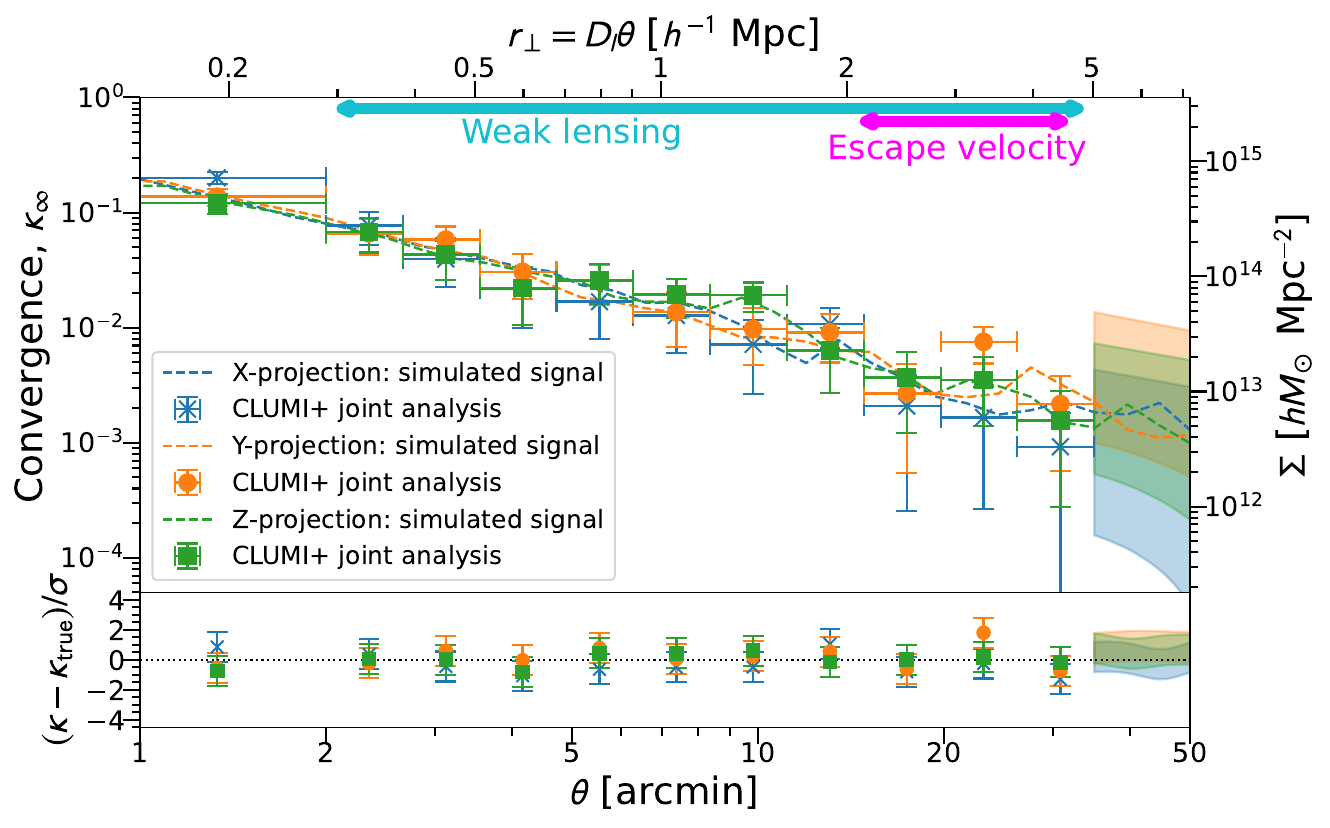}
   \includegraphics[width=0.96\columnwidth,angle=0,clip]{\FIG/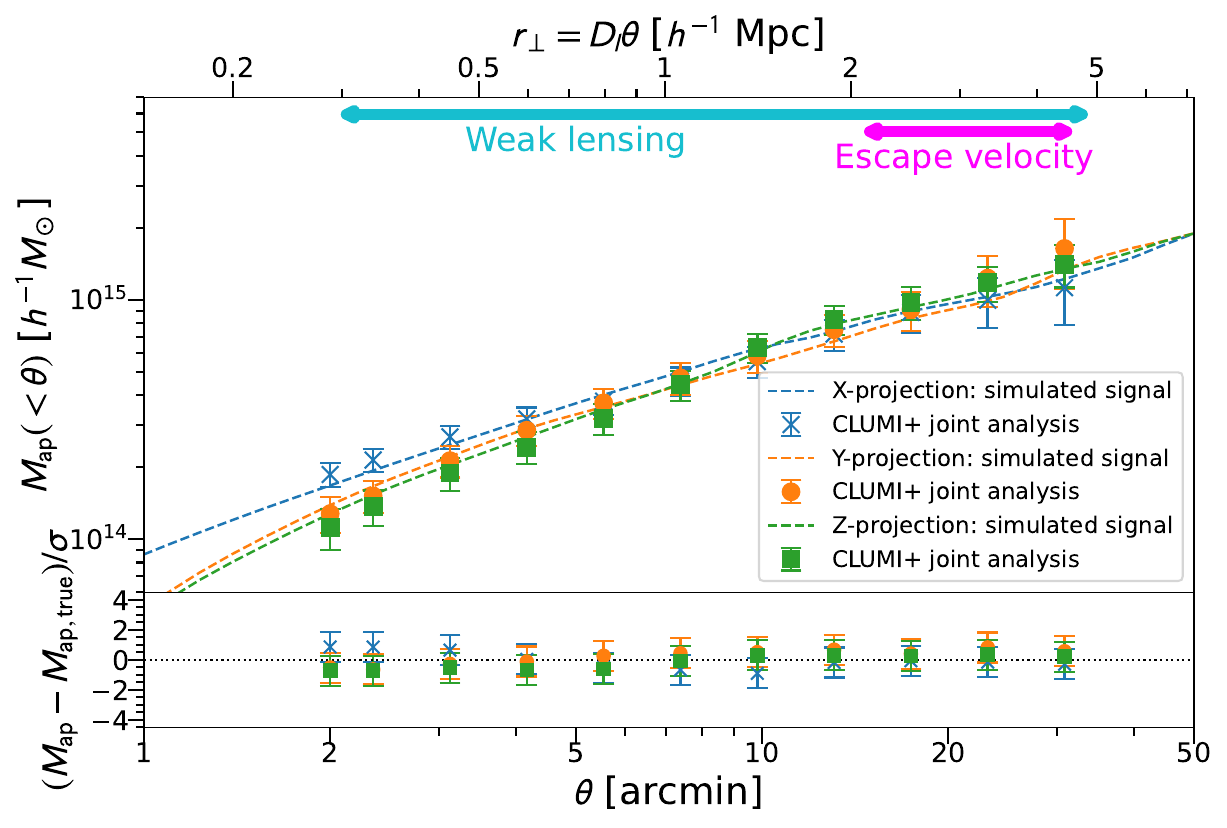}
  \end{center}
\caption{Projected mass distributions of the TNG300-1 cluster along the $X$ (blue), $Y$ (orange), and $Z$ (green) directions. The left and right panels show the lensing convergence, $\kappa_\infty(\theta) = \Sigma(r_\perp)/\SigmaCritInf$, and the cumulative aperture mass, $\Map(<\theta)=\pi r_\perp^2\overline{\Sigma}(<r_\perp)$, respectively.  In the top panels, colored symbols with error bars represent the piecewise-defined $\kappa_\infty(\theta)$ profiles for each projection (Figure~\ref{fig:tng_kappa_pdf}), reconstructed using \clumiplus from a joint analysis of synthetic weak-lensing and phase-space observations (Figures~\ref{fig:tng_mock_wl} and \ref{fig:tng_mock_vesc}). Colored dashed lines show the corresponding noise-free input profiles. The colored shaded regions in the top left panel show the marginalized $1\sigma$ confidence intervals for the external convergence term. The bottom panels display the fractional deviations of the reconstructed profiles from the true profiles, normalized by the $1\sigma$ uncertainties. The cyan and magenta horizontal bars at the top of each top panel indicate the radial ranges where weak-lensing and escape velocity constraints are used, respectively.
\label{fig:tng_kcomp}
}
\end{figure*} 

\begin{figure}[htb!] 
 \begin{center}
  \includegraphics[width=0.9\columnwidth,angle=0,clip]{\FIG/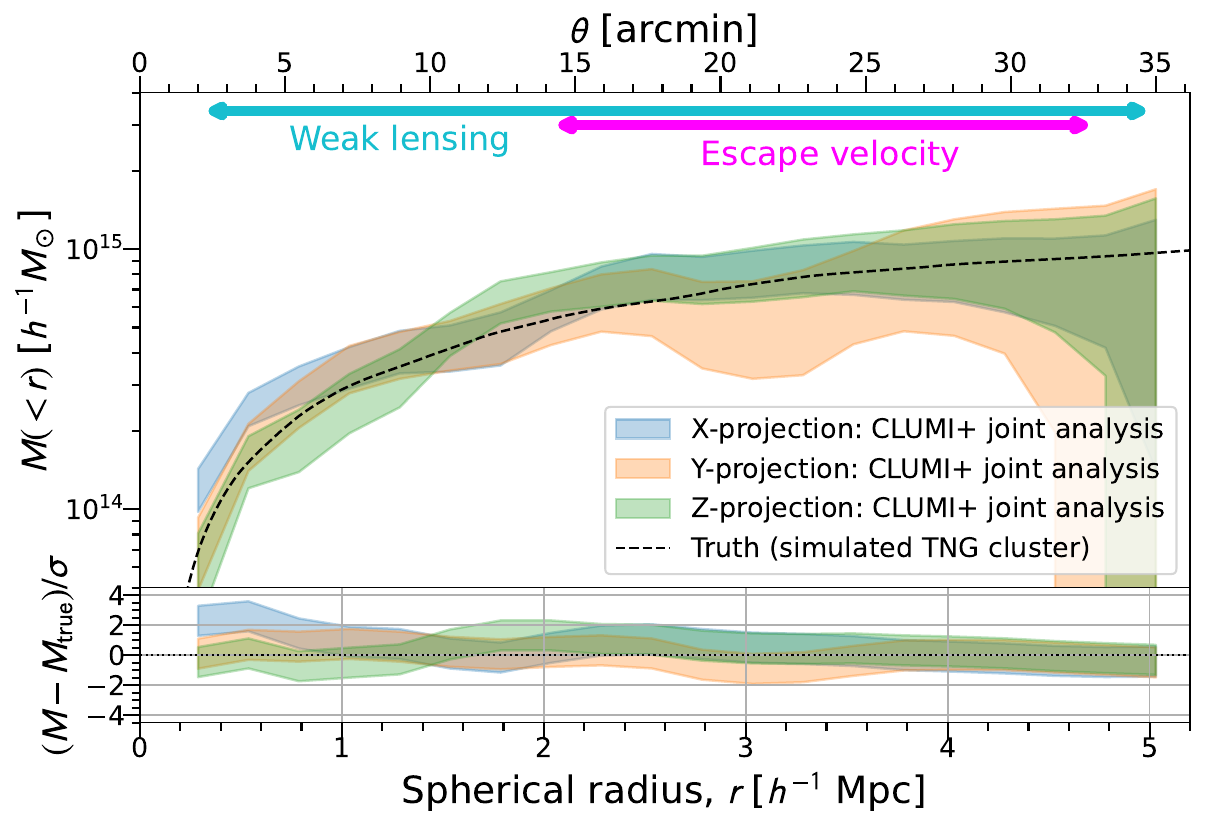}
 \end{center}
\caption{Spherically enclosed total mass profile $M(<r)$ of the TNG300-1 cluster inferred from a joint analysis of synthetic weak-lensing and phase-space observations along three orthogonal projections. In the top panel, the colored shaded regions represent the marginalized $1\sigma$ confidence intervals for each projection, and the black dashed line shows the true $M(<r)$ profile from the simulation. The bottom panel shows the fractional deviations of the recovered profiles from the true profile, normalized by the $1\sigma$ uncertainties. 
\label{fig:tng_m3dcomp}}
\end{figure} 

The results of our validation are summarized in Figures~\ref{fig:tng_kappa_pdf}--\ref{fig:tng_m3dcomp}, which assess the statistical performance and reconstruction fidelity of the \clumiplus algorithm. 

Figure~\ref{fig:tng_kappa_pdf} presents the 1D marginalized posterior distributions of the piecewise-constant convergence parameters, $\bs =\left\{ \overline{\kappa}_{\infty,\mathrm{min}}, \kappa_{\infty,i}\right\}_{i=1}^{N}$ for the $X$-projection. The left panel displays the posterior distributions of the log-transformed parameters, $P(\log_{10} \kappa_\infty)$. The right panel presents the same distributions expressed as fractional deviations from the true values, defined as $\Delta = (\kappa_\infty -\kappa_{\infty,\mathrm{true}})/\sigma$, where $\sigma$ is the standard deviation of each posterior. 

Across all bins, the posteriors are clearly single peaked, indicating that the mass-sheet degeneracy is effectively broken by the joint use of shear and magnification constraints \citep[see also Section~4.4 of][]{Umetsu+2011}. Parameters constrained with higher precision (i.e., larger values of $\kappa_\infty/\sigma$) exhibit symmetric, approximately Gaussian distributions. In contrast, posteriors in low-density bins tend to be positively skewed owing to truncation at the prior boundary, $\kappa_\infty > 0$, and are better described by lognormal-like distributions. In all bins, the marginal biweight center estimates (vertical dashed lines in the right panel) agree with the true values ($\Delta=0$) to within approximately $1\sigma$.

Figure~\ref{fig:tng_triangle} presents the 1D and 2D marginalized posterior distributions for a subset of model parameters, $(\mathcal{G}, q, \kappa_{\infty,\mathrm{ext}})$, highlighting their uncertainties and degeneracies. 
In the geometric interpretation of the caustic technique, high values of the depletion factor $\mathcal{G}\simgt 10$, as seen in the $X$- and $Z$-projections, correspond to strongly radial orbital anisotropy. Meanwhile, the posterior of the power-law slope $q$ closely tracks the adopted uniform prior across all projections, indicating that the synthetic data place only weak constraints on its value. To avoid bias in the inference, \clumiplus adopts an uninformative uniform prior, $q\in [0,2]$. Informative constraints on $q$ will likely require an ensemble cluster analysis of stacked joint data sets.
A clear positive correlation is observed between $\mathcal{G}$ and $\kappa_{\infty,\mathrm{ext}}$ in all three projections, indicating that larger values of the depletion factor are compensated by higher normalization of the external convergence term. This reflects the joint role of these parameters in determining the outer gravitational potential.

Figure~\ref{fig:tng_kcomp} displays the reconstructed projected mass distributions of the TNG300-1 cluster for all three projections. The results are presented in terms of the convergence profile, $\kappa_\infty(\theta) = \Sigma(r_\perp)/\SigmaCritInf$ (left panel), and the aperture-mass profile, $\Map(<\theta) = \pi r_\perp^2 \overline{\Sigma}(<r_\perp)$ (right panel). The corresponding true input profiles are shown for comparison. Across all projections, the reconstructions exhibit excellent agreement with the true profiles, indicating that \clumiplus accurately recovers the projected mass distribution---including the regime governed by the external convergence term---within statistical uncertainties.

Similarly, Figure~\ref{fig:tng_m3dcomp} compares the spherically enclosed total mass profiles, $M(<r)$, derived from the inferred model with the true $M(<r)$ profile of the TNG300-1 cluster from the simulation. The agreement is excellent, demonstrating that \clumiplus robustly recovers the cluster mass distribution over a wide radial range. The agreement remains excellent for the $Y$-projection, even though the caustic amplitude $\mathcal{A}(r_\perp)$ is inflated due to infalling substructures along the line of sight (Figures~\ref{fig:tng_mock_vesc} and \ref{fig:tng_caustic}). This highlights the robustness of \clumiplus to projection effects, as further discussed in Section~\ref{subsec:gains}.
Within the halo region, approximately $r \lesssim R_{200\mathrm{c}} = 1.07\Mpch$, variations among the inferred $M(<r)$ profiles across different projections reflect the impact of triaxiality \citep{Becker+Kravtsov2011,Gruen2015}. In contrast, the recovery is highly consistent in the outer and infall regions, where escape velocity information is incorporated under the assumption of spherical symmetry---validating the effectiveness of the \clumiplus algorithm in jointly constraining the mass distribution using multiprobe data.

Although the 3D mass distribution in the infall regime is generally anisotropic, the degree of anisotropy in the density field is expected to be substantially lower than in the inner halo, where triaxiality emerges from highly nonlinear structure formation \citep{Jing+Suto2002}. This reduced anisotropy likely underlies the success of spherical modeling of the gravitational potential in producing unbiased mass reconstructions at large radii.

Taken together, these findings validate the robustness of the \clumiplus framework in recovering both the projected (2D) and deprojected (3D) mass profiles of a realistic simulated galaxy cluster under synthetic observational conditions.

\begin{figure}[htb!] 
 \begin{center}
  \includegraphics[width=0.9\columnwidth,angle=0,clip]{\FIG/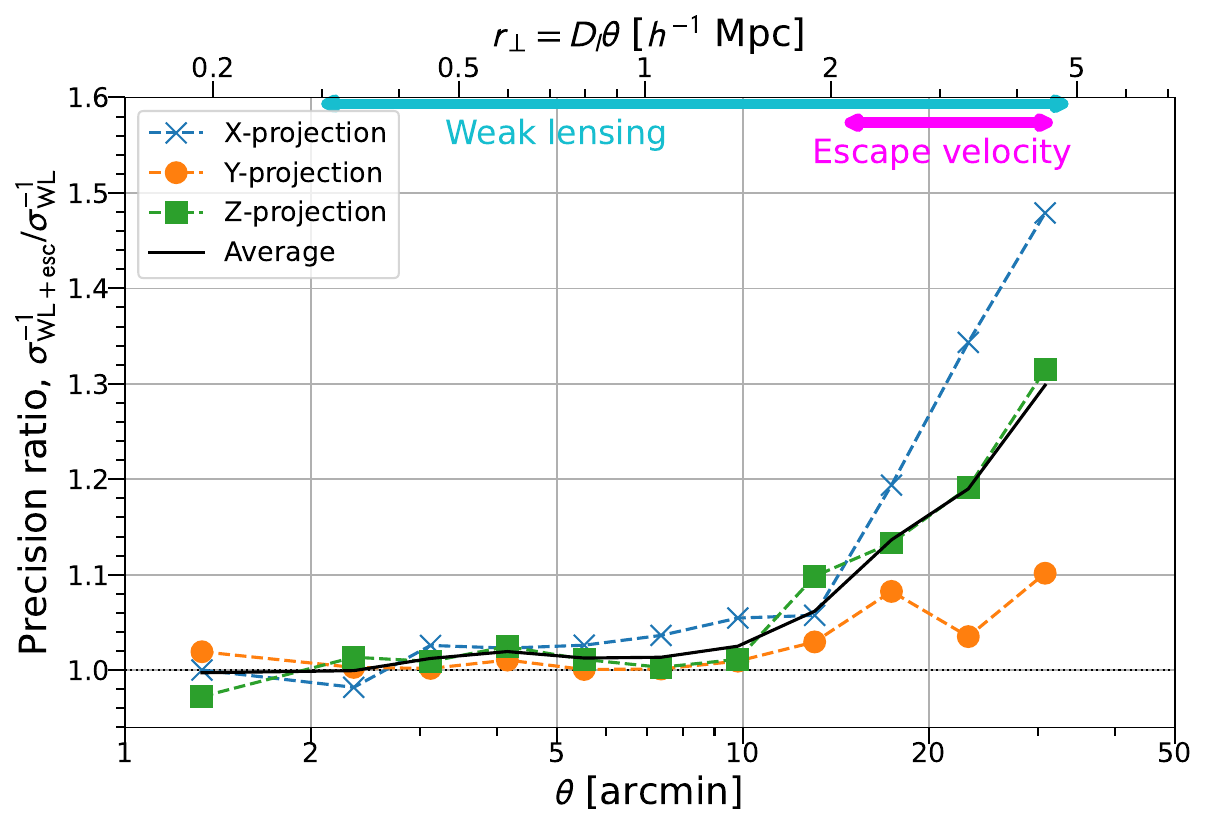}
 \end{center}
\caption{Precision gain in the piecewise-constant convergence parameters inferred from synthetic observations of the TNG300-1 cluster. We compare constraints from weak lensing alone (Figure~\ref{fig:tng_mock_wl}) with those from the joint analysis incorporating projected escape velocity measurements (Figure~\ref{fig:tng_mock_vesc}). The plotted quantity is the ratio of inverse uncertainties, $\sigma_\mathrm{WL+esc}^{-1} / \sigma_\mathrm{WL}^{-1}$, as a function of cluster-centric radius $\theta$. Results are shown for three orthogonal projections: $X$ (blue crosses), $Y$ (orange circles), and $Z$ (green squares). The solid black line denotes the mean precision gain across all projections.
\label{fig:tng_kprec}}
\end{figure}

\subsection{Precision Gain from Joint \clumiplus Analysis}
\label{subsec:gains}

One of the principal advantages of \clumiplus is its improved precision in mass reconstruction, achieved by incorporating escape velocity information from projected phase-space observations. To demonstrate this, we performed a lensing-only \textsc{clumi} analysis of the synthetic weak-lensing data, $\{\langle g_{+,i}\rangle, \langle n_{\mu,i}\rangle\}_{i=1}^{\NWL}$, for the TNG300-1 cluster and quantified the resulting gains in the precision of the piecewise-defined convergence parameters, $\bs=\{\kappa_{\infty,\mathrm{min}},\kappa_{\infty,i}\}_{i=1}^{N}$.  

We define the precision gain as the ratio of inverse posterior uncertainties, $\sigma^{-1}_\mathrm{WL+esc} / \sigma^{-1}_\mathrm{WL}$, where $\sigma_\mathrm{WL+esc}$ and $\sigma_\mathrm{WL}$ denote the marginalized posterior standard deviations from the joint \clumiplus and lensing-only \clumi analyses, respectively. This metric quantifies the statistical improvement achieved by incorporating projected escape velocity measurements into the mass reconstruction.

The results are shown in Figure~\ref{fig:tng_kprec}. On average, we find a $10\percent$--$30\percent$ improvement in precision in the outer regions where the weak-lensing and phase-space data sets overlap ($r_\perp > 2\Mpch$). Notably, within this overlapping regime, the level of precision gain increases with cluster-centric radius, reflecting the growing relative constraining power of the escape velocity measurements at larger scales. This trend suggests that broader radial overlap between weak-lensing and spectroscopic data will yield even greater precision gains, underscoring the potential of the \clumiplus framework for future wide-field survey applications.

We also find that the magnitude of the precision gain is sensitive to projection effects, which can induce discrepancies between weak-lensing and escape velocity constraints. In the $Y$-projection, for example, two massive substructures infalling along the line of sight and located outside the main halo produce elevated caustic amplitudes (see Section~\ref{subsubsec:tng_mock_vesc}). In particular, a local enhancement in the caustic amplitude around $r_\perp \sim 2.3\Mpch$ is not well described by the smoothly varying gravitational potential $\Phi_\mathrm{N}(r)$ assumed in the model. As a result, the precision gain is locally suppressed in radial regions where the lensing and dynamical constraints are in tension (compare Figures~\ref{fig:tng_mock_vesc} and \ref{fig:tng_kprec}), illustrating the impact of projection effects on the fidelity of mass reconstruction.

We emphasize that chance alignments involving line-of-sight infall or cluster--cluster interactions may reduce the statistical gains achievable from joint modeling, but they do not introduce bias into the reconstructed mass profile. On the contrary, localized discrepancies between lensing and dynamical observables can serve as diagnostics for identifying projection effects associated with line-of-sight structure. The \clumiplus framework remains robust in such scenarios, due to its reliance on complementary, physically distinct probes of the gravitational potential.

\subsection{Accuracy and Bias in Profile Reconstruction}
\label{subsec:accuracy}

To complement our analysis of statistical precision in Section~\ref{subsec:gains}, we assess the \textit{accuracy} and potential \textit{bias} of the reconstructed binned convergence profiles by comparing them directly to the known ground truth. 

To summarize the performance across the full profile, we define global accuracy and precision metrics:\footnote{We avoid defining accuracy as the unweighted average of fractional deviations, i.e., $(N+1)^{-1} \sum_i |\kappa_i-\kappa_{\mathrm{true},i}|/\kappa_{\mathrm{true},i}$, where $(N+1)$ is the total number of bins including the innermost bin interior to the minimum radius $\thetamin$. This metric disproportionately amplifies deviations in bins where the signal is intrinsically small and the measurement uncertainty is large, regardless of their physical significance. In contrast, we obtain a more balanced and physically meaningful measure of reconstruction fidelity as defined in Equation~(\ref{eq:metric}).}
\begin{equation}
\label{eq:metric}
\begin{aligned}
\overline{\delta} &= \frac{\sum_i |\kappa_i - \kappa_{\mathrm{true},i}|}{\sum_i \kappa_{\mathrm{true},i}}, \\
\overline{\Delta} &= \frac{\sum_i \sigma_{\kappa,i}}{\sum_i \kappa_{\mathrm{true},i}},
\end{aligned}
\end{equation}
where $\kappa_i$ and $\sigma_{\kappa,i}$ denote the inferred convergence and its statistical uncertainty, respectively, and $\kappa_{\mathrm{true},i}$ is the true convergence in bin $i$. These scalar quantities quantify the typical per-bin fractional deviation and uncertainty, respectively.

We apply this diagnostic to compare the \clumiplus and \clumi reconstructions of our TNG300-1 cluster along three orthogonal projections.  The resulting accuracy metrics are summarized as follows:
\begin{itemize}
    \item $X$-projection: $\overline{\delta} = 0.171$ (\clumiplus), $0.192$ (\clumi);
    \item $Y$-projection: $\overline{\delta} = 0.133$ (\clumiplus), $0.143$ (\clumi);
    \item $Z$-projection: $\overline{\delta} = 0.116$ (\clumiplus), $0.129$ (\clumi).
\end{itemize}
Thus, \clumiplus consistently achieves lower absolute fractional deviations than \clumi. In all cases, the reconstruction accuracy is better than the statistical precision ($\overline{\Delta} \sim 27$--$31\%$), i.e., $\overline{\delta} < \overline{\Delta}$, indicating that the inferred profiles are statistically consistent with the ground truth and show no evidence of systematic bias.  These results demonstrate that the inclusion of escape velocity information in \clumiplus improves both the accuracy and overall fidelity of the mass reconstruction relative to lensing-only constraints.

\section{Application to Cluster Observations: A2261}
\label{sec:a2261}

In this section, we apply the \clumiplus algorithm to multiprobe observations of the galaxy cluster A2261 (hereafter A2261) at $z = 0.225$, a well-studied, massive system targeted by both the CLASH program \citep{Postman2012clash} and the Hectospec Cluster Survey \citep[HeCS;][]{Rines2013HeCS}. Specifically, we combine the CLASH weak- and strong-lensing data products from \citet{Umetsu2016clash} with projected escape velocity constraints derived from HeCS spectroscopic observations presented by \citet{pizzardo2020} (see also \citealt{Rines2013HeCS}). We note that strong-lensing constraints were not included in the simulation-based validation presented in Section~\ref{sec:tng}, but are incorporated here for completeness in the analysis of real observational data.

A2261 is part of the X-ray-selected CLASH subsample, which comprises 20 clusters with X-ray temperatures exceeding 5~keV and exhibiting a high degree of regularity in their X-ray morphology. The cluster has an effective Einstein radius of $\Rein = 23.1\arcsec$ for sources at $\zs = 2$ \citep{Zitrin2015clash,Umetsu2016clash}. Its mass has been estimated as $M_{500\mathrm{c}} = (11.0 \pm 2.1) \times 10^{14}\Msunh$ from a joint weak- and strong-lensing analysis assuming a spherical NFW profile \citep{Umetsu2016clash} (see also \citealt{Coe2012}). A2261 is also identified as a massive SZ cluster detected by the \Planck satellite, with a mass estimate of $M_{500\mathrm{c,SZ}} = (5.5 \pm 0.2) \times 10^{14}\Msunh$ based on the \Planck SZ scaling relation \citep{Planck2014XX}, which is calibrated using X-ray observations and hydrodynamical simulations under the assumption of hydrostatic equilibrium.

The substantial discrepancy between the lensing and SZ mass estimates suggests that A2261 is likely not in equilibrium. This highlights the importance of employing multiprobe approaches that do not rely on equilibrium assumptions for accurate mass reconstruction.

\subsection{Data Summary}
\label{subsec:a2261_data}

\subsubsection{CLASH Lensing Data}
\label{subsubsec:a2261_lens}

The weak- and strong-lensing data used in this analysis were originally presented by \citet{Umetsu2016clash} as part of the CLASH program \citep{Postman2012clash}. The weak-lensing shear and magnification profiles were derived from deep Subaru/Suprime-Cam imaging in five optical bands \citep{Umetsu2014clash}, while central aperture-mass constraints were obtained from strong-lens modeling based on 16-band Hubble Space Telescope (\HST) imaging \citep{Zitrin2015clash}.

In the CLASH collaboration, mass reconstructions were performed using two independent methods: \textsc{sawlens} \citep{Merten2015clash} and \clumi \citep{Umetsu2016clash}. 
\textsc{sawlens} is a free-form, pixel-based 2D reconstruction technique that combines strong-lensing catalogs of multiply imaged systems with weak-lensing shear catalogs of background galaxies. 
In contrast, \clumi employs a likelihood-based framework formulated at the level of azimuthally averaged radial profiles (Section~\ref{subsec:model}). Despite these methodological differences, both approaches yielded consistent reconstructions of the convergence profile $\kappa_\infty(\theta)$ \citep[see Appendix~B of][]{Umetsu2016clash}.

The Subaru weak-lensing profiles, $\{\langle g_{+,i}\rangle, \langle n_{\mu,i}\rangle\}_{i=1}^{\NWL}$, were measured over the radial range $\theta \in [0.9\arcmin, 16\arcmin]$ centered on the BCG, using $\NWL = 10$ bins. The strong-lensing constraints, $\{M_{\mathrm{ap},i}\}_{i=1}^{\NSL}$, were derived in the range $\theta \in [10\arcsec, 40\arcsec]$ with a uniform sampling interval of $10\arcsec$, yielding $\NSL = 4$ constraints \citep[Table~1 of][]{Umetsu2016clash}. This sampling provides an optimal representation of the strong-lensing $\Map$ constraints for the CLASH sample, which has a median Einstein radius of $\Rein \approx 20\arcsec$ for sources at $\zs = 2$. Together, the weak+strong-lensing measurements in $N=\NWL+\NSL=14$ bins provide well-sampled constraints on the projected mass distribution from the cluster core to the outskirts. 

In Appendix~\ref{appendix:a2261_data} (Figure~\ref{fig:a2261_wl}), we present the Subaru weak-lensing profiles of A2261 analyzed in this study. The total detection $\mathrm{S/N}$ values are $16.4$ and $8.1$ for the tangential distortion and magnification bias measurements, respectively, with a combined $\mathrm{S/N}$ of $\sim 18$. The background source population used for the magnification analysis is characterized by a mean unlensed surface number density of $\overline{n}_\mu = (20.0 \pm 0.4)\,\mathrm{galaxies\,arcmin}^{-2}$ and a logarithmic count slope of $\alpha = 0.35 \pm 0.10$ \citep{Umetsu2014clash}.

\begin{figure}[htb!] 
 \begin{center}
  \includegraphics[width=\columnwidth,angle=0,clip]{\FIG/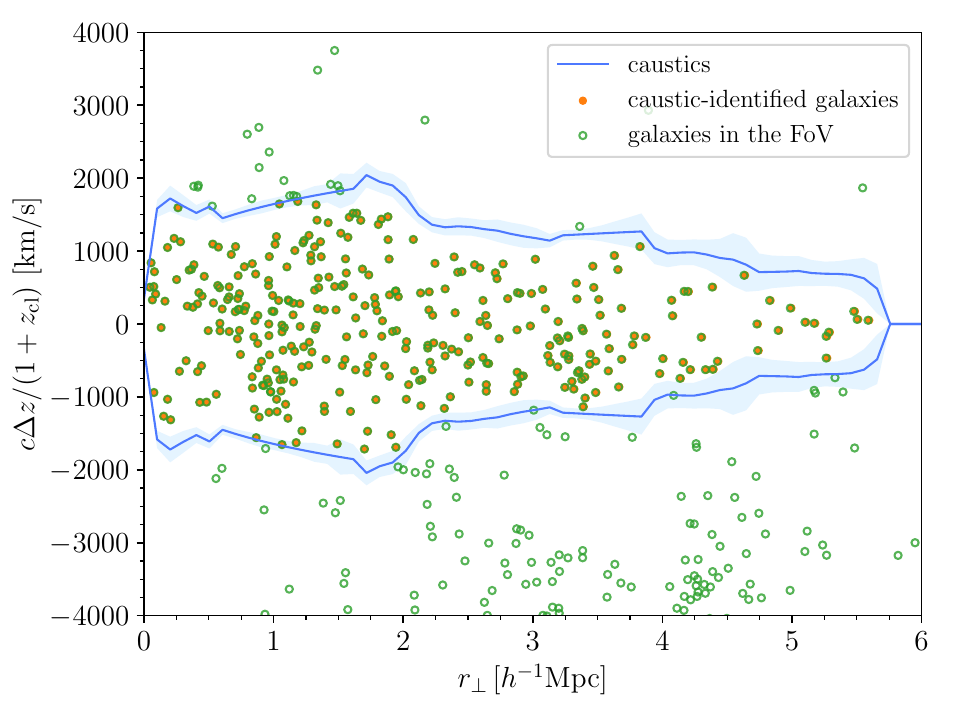}
 \end{center}
\caption{Redshift diagram of the galaxy cluster A2261 at $z_\mathrm{cl} = 0.225$ based on spectroscopic observations from HeCS. The line-of-sight velocity of galaxies, $v_\mathrm{los} = c\Delta z / (1 + z_\mathrm{cl})$, relative to the BCG is plotted as a function of projected cluster-centric radius, $r_\perp$. The blue curves indicate the caustic boundaries derived using the caustic technique. Green circles denote all galaxies within the field of view, while filled orange circles represent galaxies located within the caustic boundaries, corresponding to candidate bound or infalling systems.
\label{fig:a2261_rvlos}}
\end{figure}

\subsubsection{Hectospec Spectroscopic Data}
\label{subsec:a2261_caustic}

For the phase-space analysis, we utilize galaxy redshifts from HeCS \citep{Rines2013HeCS,Sohn2021hecsdr1}, which provides dense spectroscopic sampling of galaxies around galaxy clusters, ideally suited for applications of the caustic technique. HeCS leverages archival data from the Sloan Digital Sky Survey (SDSS) and the \ROSAT All-Sky Survey (RASS). Specifically, X-ray cluster catalogs derived from RASS were used to define flux-limited cluster samples, which were subsequently matched to the imaging footprint of SDSS Data Release 6 \citep[DR6;][]{Adelman_McCarthy_2006}. The multiband SDSS photometry enabled the identification of candidate cluster members using the red-sequence technique. 
To obtain reliable redshift measurements beyond $z \sim 0.1$ where SDSS spectroscopic sampling becomes sparse, HeCS employed the Hectospec multiobject spectrograph \citep{Fabricant2005Hecto} on the MMT 6.5~m telescope to obtain dedicated spectroscopic observations of cluster-associated galaxies.

\citet{pizzardo2020} present a caustic analysis for the HeCS cluster catalog using the method implemented by \citet{Serra2011}, which refines the original algorithm developed by \citet{Diaferio1999}. The updated caustic profiles show good agreement with those reported by \citet{Rines2013HeCS}, with differences well within the $1\sigma$ uncertainties.

For A2261, a total of 335 spectroscopically confirmed galaxies within the caustic boundaries, corresponding to candidate bound or infalling systems \citep[see][]{Serra2013}, are identified within a projected radius of approximately $10$~Mpc ($\sim 7\Mpch$).
As shown in Figure~\ref{fig:a2261_rvlos}, the redshift diagram exhibits a prominent caustic structure, from which we derive the projected escape velocity profile, $\mathcal{A}(r_\perp)$, using the caustic technique. We extract projected escape velocity measurements, $\{\mathcal{A}_{i}\}_{i=1}^{\NEV}$, over the radial range $r_\perp \in [2, 5]\Mpch$, with a uniform sampling interval of $\overline{h}_r = 343\kpch$, yielding $\NEV = 8$ constraints. The resulting $\mathcal{A}(r_\perp)$ profile provides dynamical constraints on the gravitational potential at $r_\perp > \Rinf$, with a total detection $\mathrm{S/N}$ of $\sim 14$ (see Figure~\ref{fig:a2261_vesc} in Appendix~\ref{appendix:a2261_data}). This is comparable to the combined weak-lensing $\mathrm{S/N}$ of $\sim 18$ (Section~\ref{subsubsec:a2261_lens}).

\begin{figure}[htb!] 
 \begin{center}
  \includegraphics[width=\columnwidth,angle=0,clip]{\FIG/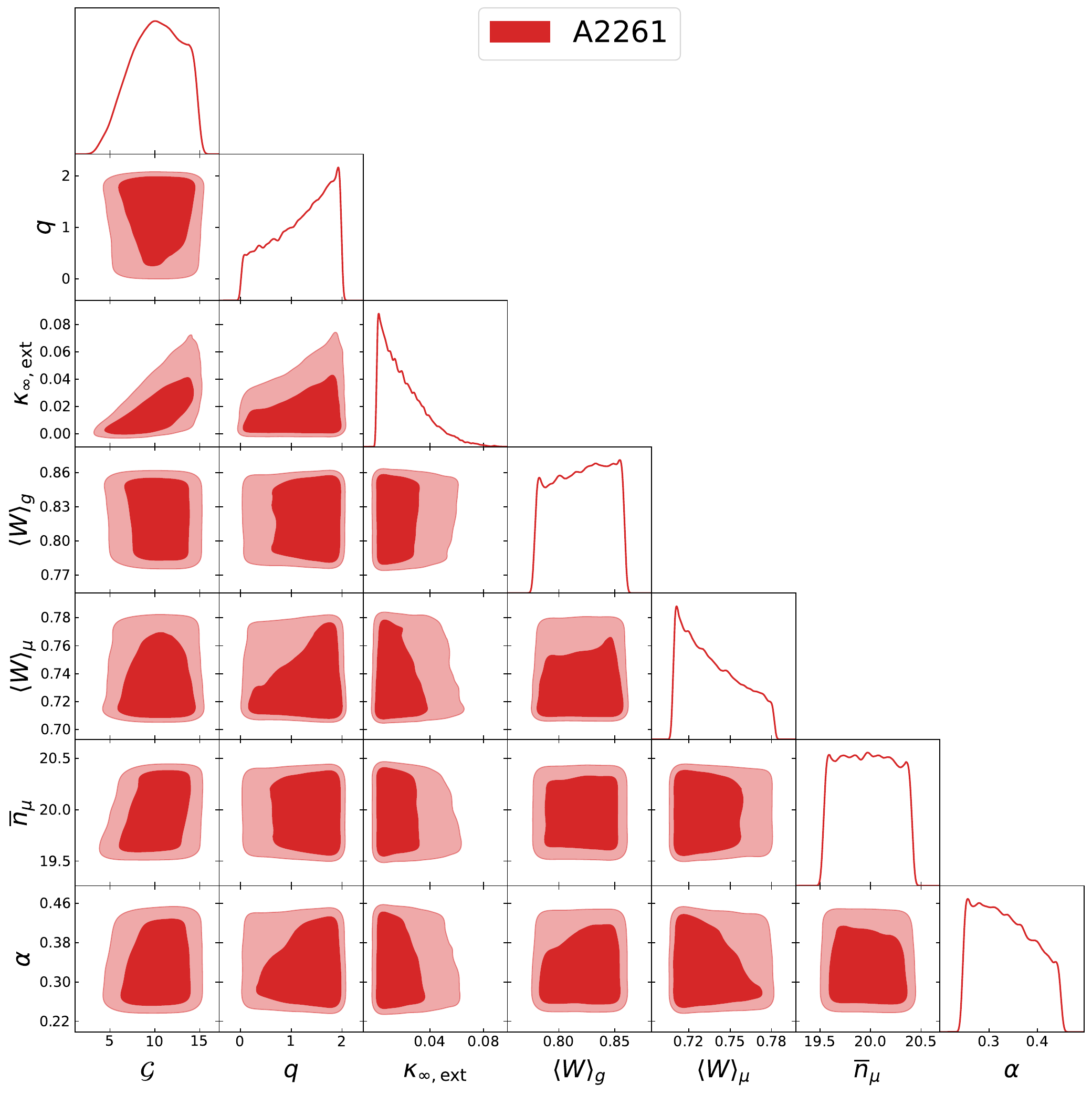}
 \end{center}
\caption{Marginalized posterior distributions (68\percent and 95\percent confidence levels) for a subset of the model parameters $(\mathcal{G}, q, \kappa_{\infty,\mathrm{ext}}, \langle W\rangle_g, \langle W\rangle_\mu, \overline{n}_\mu, \alpha)$ for A2261, inferred from a joint \clumiplus analysis of strong-lensing, weak-lensing, and phase-space observations. The diagonal panels show the 1D marginalized posterior distributions, while the off-diagonal panels display the 2D joint posterior distributions for each parameter pair.
\label{fig:a2261_triangle}
}
\end{figure}

\begin{figure*}[htb!] 
 \begin{center}
  \includegraphics[width=0.9\textwidth,angle=0,clip]{\FIG/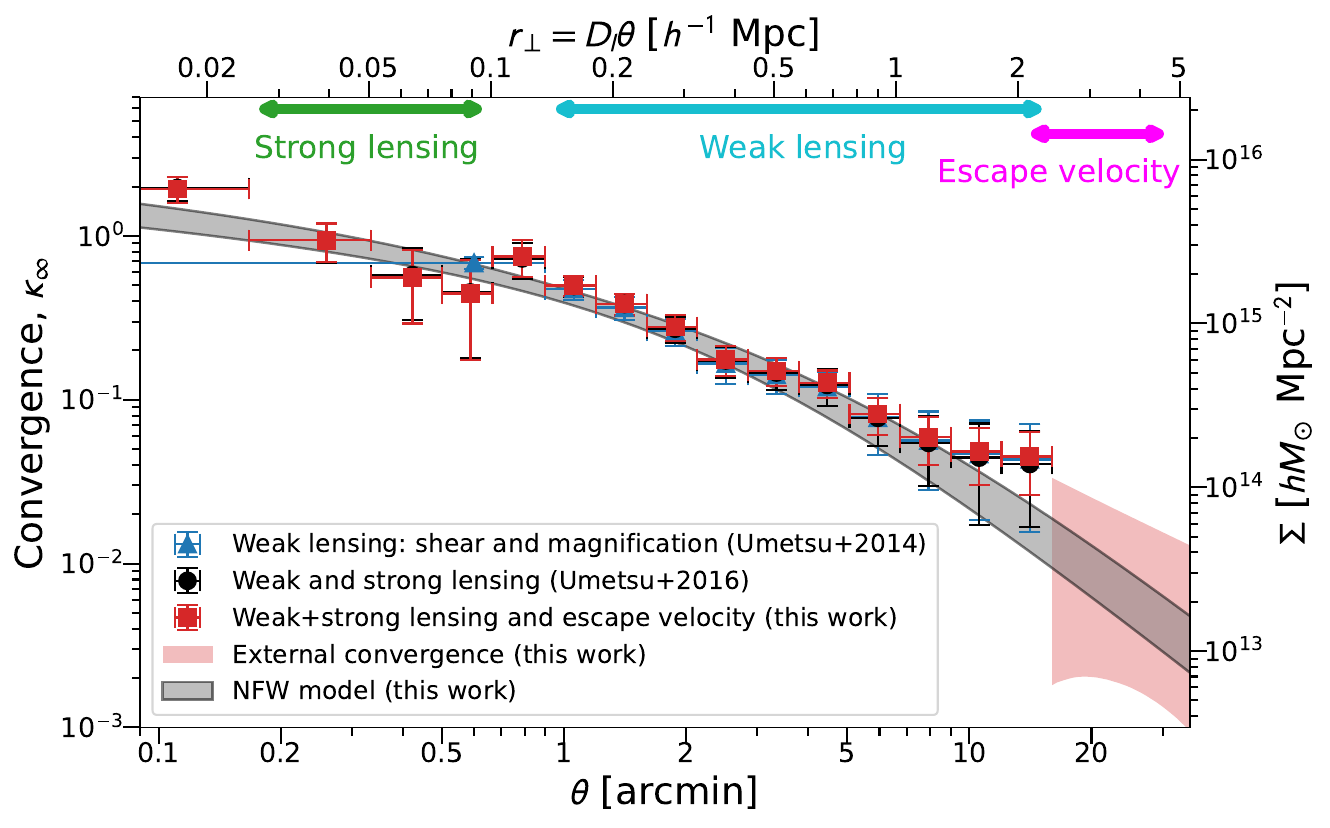}
 \end{center}
\caption{Surface mass density profile, $\kappa_\infty(\theta) = \Sigma(r_\perp) / \SigmaCritInf$, of A2261 as a function of cluster-centric radius, $\theta = r_\perp / D_l$. Red squares with error bars represent the piecewise-defined $\kappa_\infty(\theta)$ profile inferred from the joint \clumiplus analysis of strong-lensing, weak-lensing, and phase-space observations (see Figures~\ref{fig:a2261_wl} and \ref{fig:a2261_vesc}). The red shaded region indicates the marginalized $1\sigma$ uncertainty on the external convergence term. The gray shaded region shows the $1\sigma$ confidence interval from an NFW fit to the \clumiplus-derived profile (red squares). For comparison, we also show weak-lensing-only (blue triangles; \citealt{Umetsu2014clash}) and combined weak+strong-lensing (black circles; \citealt{Umetsu2016clash}) results obtained with \clumi. Horizontal bars at the top mark the radial ranges of the observational constraints: strong lensing (green), weak lensing (cyan), and escape velocity (magenta).
\label{fig:a2261_kplot}}
\end{figure*}

\begin{figure}[htb!] 
 \begin{center}
  \includegraphics[width=0.9\columnwidth,angle=0,clip]{\FIG/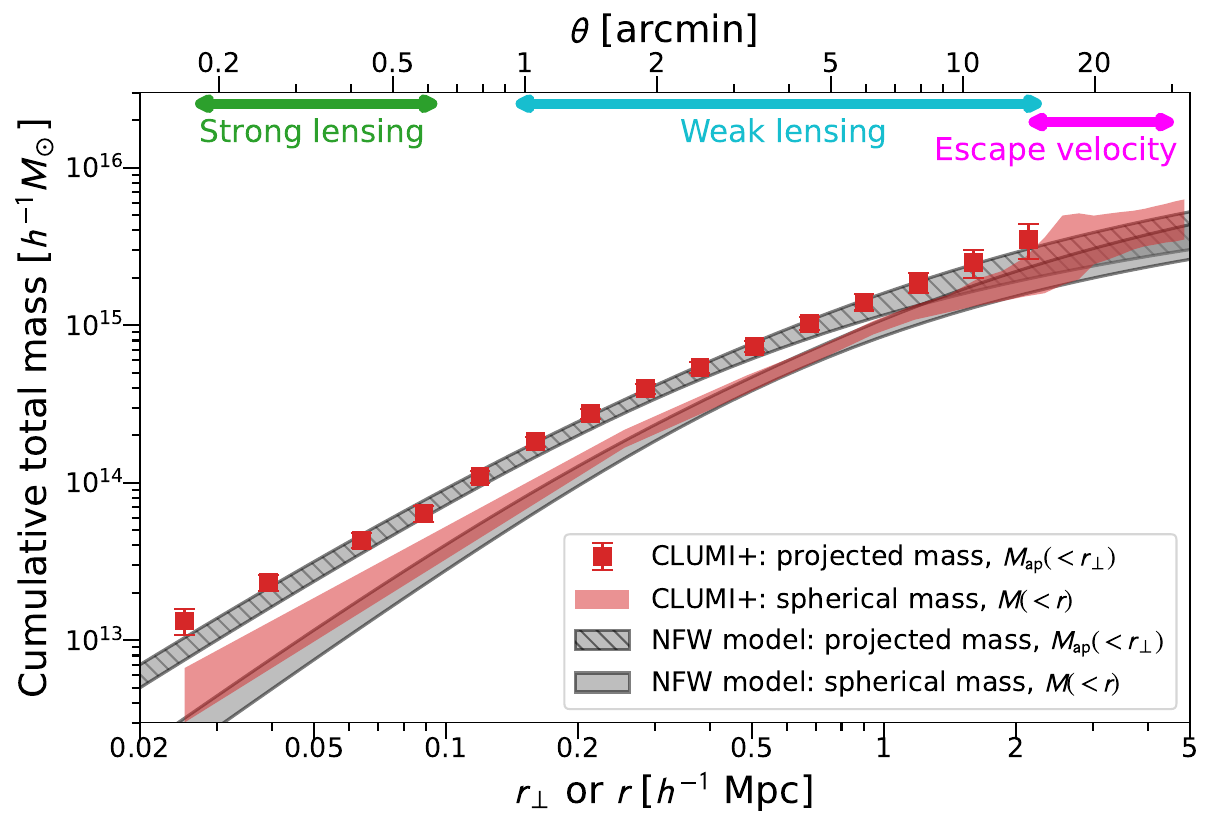}
 \end{center}
\caption{Cumulative total mass distributions of A2261 inferred from the joint \clumiplus analysis. Red squares with error bars show the projected enclosed mass profile, $\Map(<r_\perp)$, and the red shaded region shows the spherically enclosed (3D) mass profile, $M(<r)$. The gray shaded region and the gray hatched region represent the marginalized $1\sigma$ confidence intervals for the spherical and projected mass profiles, respectively, obtained from an NFW fit to the piecewise-defined convergence profile shown in Figure~\ref{fig:a2261_kplot}.
\label{fig:a2261_mplot}
}
\end{figure}

\subsection{Results and Comparison}
\label{subsec:a2261_results}

We perform a joint likelihood analysis of the CLASH lensing data and the projected escape velocity measurements from HeCS for A2261 using the \clumiplus framework. As in our validation study (Section~\ref{sec:tng}), the mass distribution is modeled with a piecewise-defined convergence profile within the lensing aperture ($\thetamax=16\arcmin$), transitioning to a projected power-law form that describes the external mass distribution beyond it. The full \clumiplus model, $\bp=\{\bm,\bc\}$, comprises $N+8=22$ parameters: the hybrid mass model $\bm = \{\bs, \mathcal{G}, q, \kappa_{\infty,\mathrm{ext}}\}$ specified by $N + 4$ parameters and four calibration nuisance parameters $\bc$ (see Section~\ref{sec:clumiplus}).

Figure~\ref{fig:a2261_triangle} presents the marginalized 1D and 2D posterior distributions for a subset of the model parameters, $(\mathcal{G}, q, \kappa_{\infty,\mathrm{ext}}, \langle W\rangle_g, \langle W\rangle_\mu, \overline{n}_\mu, \alpha)$, excluding the piecewise-defined convergence parameters $\bs$, to illustrate parameter uncertainties and degeneracies. 

Figure~\ref{fig:a2261_kplot} shows the reconstructed convergence profile, $\kappa_\infty(\theta) = \Sigma(r_\perp) / \SigmaCritInf$, with marginalized $1\sigma$ confidence intervals, where $\SigmaCritInf(\zl) = 3.42\,h \times 10^{15}\,M_\odot\,\mathrm{Mpc}^{-2}$ is the far-background critical surface mass density. The \clumiplus-derived profile is in good agreement with previous CLASH lensing-only reconstructions obtained with \clumi \citep{Umetsu2014clash,Umetsu2016clash}, while achieving improved precision in the outer regions owing to the inclusion of projected escape velocity constraints (see Section~\ref{subsec:discussion} for details).

To characterize the mass distribution and derive 3D halo parameters, we perform a spherical NFW fit to the reconstructed piecewise-defined $\kappa_\infty(\theta)$ profile, $\bs = \{\kappa_{\infty,\mathrm{min}}, \kappa_{\infty,i}\}_{i=1}^{N}$, using the full covariance matrix $(C)_{ij}$:
\begin{equation}
 C = C^\mathrm{stat} + C^\mathrm{lss} + C^\mathrm{int},
\end{equation}
where $C^\mathrm{stat}$ is the posterior statistical covariance matrix  from \clumiplus (Equation~\ref{eq:cstat}), $C^\mathrm{lss}$ is the cosmic noise covariance due to uncorrelated large-scale structure projected along the line of sight, and $C^\mathrm{int}$ accounts for statistical fluctuations in the lensing signal at fixed halo mass arising from intrinsic variations and projection effects \citep{Umetsu2020rev}. The elements of $C^\mathrm{lss}$ and $C^\mathrm{int}$ are computed following the prescriptions of \citet{Umetsu2016clash}.

We adopt uninformative uniform priors on the NFW parameters, $\log_{10}(M_{200\mathrm{c}}/h^{-1}\Msun) \in [14, 16]$ and $\log_{10}c_{200} \in [-1, 1]$, consistent with \citet{Umetsu2016clash}. From the marginalized posterior distributions, we obtain constraints of $M_{200\mathrm{c}} = (17.1 \pm 4.1)\times 10^{14}\Msunh$ and $c_\mathrm{200c} = 3.43 \pm 0.89$ for A2261, in agreement with the weak+strong-lensing results of \citet{Umetsu2016clash} based on \clumi. The corresponding overdensity radius is $R_{200\mathrm{c}} = (1.80 \pm 0.14) \Mpch$.
From the posterior samples, we also derive $M_{500\mathrm{c}}=(11.3 \pm 2.3)\times 10^{14}\Msunh$, which again points to a large discrepancy with the \Planck mass estimate, $M_{500\mathrm{c,SZ}} = (5.5 \pm 0.2) \times 10^{14}\Msunh$. The best-fit NFW profile is compared with the \clumiplus reconstruction in Figure~\ref{fig:a2261_kplot}.

Figure~\ref{fig:a2261_mplot} presents the cumulative projected and spherical mass profiles, $\Map(<r_\perp)$ and $M(<r)$, respectively, derived from the joint \clumiplus analysis. The inferred profiles from the flexible \clumiplus model $\bm$ show good agreement with the best-fit NFW predictions across a broad radial range. A modest excess is observed in the innermost region probed by \HST strong lensing, which may reflect the baryonic contribution of the BCG not captured by the NFW model.

The fact that an NFW profile provides a good fit to the lensing signal of A2261 does not imply that the cluster is relaxed. Rather, it reflects the empirical finding that NFW profiles often describe the projected mass distributions of both relaxed and unrelaxed clusters \citep{Umetsu2020rev}. This is consistent with expectations for halos dominated by collisionless CDM \citep{Child2018cm}, where the dark matter responds to dynamical disturbances on timescales of $\sim 1$~Gyr---shorter than the gas relaxation timescales \citep{Ricker+Sarazin2001,Umetsu2010}.

\subsection{Discussion}
\label{subsec:discussion}

The application of \clumiplus to A2261 demonstrates the effectiveness of combining weak and strong lensing with projected phase-space information for cluster mass reconstruction. The resulting mass profile is consistent with previous CLASH analyses while achieving improved precision in the outer regions, where lensing constraints alone become increasingly limited. This case study highlights the practical utility of \clumiplus for joint modeling based on complementary observables.

Since the Subaru lensing aperture is limited to $D_l \thetamax \approx 2.4\Mpch$, which is comparable to the adopted boundary of the infall region, $\Rinf = 2\Mpch$, there is minimal radial overlap between the lensing and dynamical data sets. Nevertheless, the addition of projected escape velocity measurements, extending out to $r_\perp \sim 4.8\Mpch$, yields improvements in the precision of the piecewise-defined convergence parameters, $\bs = \{\kappa_{\infty,\mathrm{min}}, \kappa_{\infty,i}\}_{i=1}^{N}$, particularly at large radii. In the outer regions ($r_\perp \gtrsim 1\Mpch$), we find precision gains of $\sim 20\percent$ (Figure~\ref{fig:a2261_kplot}). Notably, the lensing and phase-space constraints exhibit good mutual consistency with the joint solution inferred from \clumiplus (see Figures~\ref{fig:a2261_wl} and \ref{fig:a2261_vesc} in Appendix~\ref{appendix:a2261_data}), consistent with the validation results from the IllustrisTNG cluster simulations.

However, we note a potential discrepancy between the observed gradient of the caustic amplitude $A(r_\perp)$ and the \clumiplus inference in A2261 (Figure~\ref{fig:a2261_vesc}), similar to that seen in the $Y$-projection of the TNG300-1 cluster (Figure~\ref{fig:tng_mock_vesc}). In A2261, the observed $A(r_\perp)$ increases with radius beyond $r_\perp\sim 1\Mpch$, reaching a peak near $r_\perp \sim 1.8\Mpch$ followed by a sharp decline around $r_\perp \sim 2\Mpch$ (Figure~\ref{fig:a2261_rvlos}), a trend at odds with theoretical expectations, since the escape velocity should decline monotonically with radius. This behavior is best explained as a projection effect from multiple substructures located in the outer infall region whose infall velocities---both approaching and receding relative to the observer---are aligned along the line of sight. Such alignments, particularly in the cluster outskirts, where tidal forces are weaker and substructures are more likely to remain intact, can locally thicken the phase-space envelope, inflating $A(r_\perp)$ and producing a steep outward decline unrelated to the true gravitational potential (see Section~\ref{subsec:tng_results}). 
Supporting this interpretation, wide-field weak-lensing studies of A2261 have identified several significant mass substructures in the outskirts \citep[e.g.,][]{Okabe2010wl, Coe2012, Umetsu2014clash, Dutta2025}. The \clumi and \clumiplus  reconstructions also reveal a localized surface mass excess in the outermost radial bins relative to the best-fit NFW profile, while the external convergence term inferred from the joint analysis remains consistent with the large-scale NFW trend (Figure~\ref{fig:a2261_kplot}).

Although the inclusion of projected escape velocity constraints yields clear improvements in the precision of the mass profile at large radii (Figure~\ref{fig:a2261_kplot}), we find that the inferred NFW parameters, $M_{200\mathrm{c}}$ and $c_{200\mathrm{c}}$, are statistically consistent in precision values with those obtained by \citet{Umetsu2016clash} using strong and weak lensing alone. This behavior is expected, since the NFW parameters are primarily sensitive to the density structure within the halo, whereas the precision gains from \clumiplus occur predominantly in the outer and infall regions. These findings underscore the complementarity of lensing and phase-space information and highlight the specific utility of \clumiplus in probing outer density structure beyond the virial radius.

In particular, such precision gains in the outskirts are especially valuable for constraining the steepening feature in the radial density profile associated with the splashback radius, $R_{\mathrm{sp}}$, as probed by cluster--galaxy weak lensing \citep[e.g.,][]{Umetsu+Diemer2017,Chang2018sp}. The splashback radius marks a physically motivated boundary of the cluster halo, corresponding to the location where recently accreted matter reaches its first apocenter, giving rise to a sharp drop in the density profile \citep{Diemer+Kravtsov2014,More2015splash}. Resolving this feature with high fidelity requires precise measurements over a broad radial range extending from the halo region into the infall regime. In this context, \clumiplus offers a promising avenue for detecting and characterizing splashback features in ensemble cluster analyses, by harnessing the combined constraining power of stacked lensing and phase-space observables across the full dynamical range of the cluster environment.

For A2261, as well as for all three projections of the TNG300-1 cluster, we find consistently high inferred values of the caustic depletion factor in the infall region, with $\mathcal{G} \gtrsim 10$ (Figures~\ref{fig:tng_triangle} and \ref{fig:a2261_triangle}). In the geometric interpretation of the caustic technique (Section~\ref{sec:caustic}), such values correspond to strongly radial orbital anisotropy, with velocity anisotropy parameters $\beta_v \gtrsim 0.88$, 
or equivalently $\sqrt{(\langle v_\theta^2\rangle + \langle v_\phi^2\rangle) / 2\langle v_r^2\rangle} \lesssim 0.35$. 
These results may reflect genuine dynamical features in the cluster outskirts, where infall and radial streaming motions dominate the tracer population.

Alternatively, this trend may indicate a mild systematic overestimation of the 3D escape velocity, $\vesc$, at fixed projected amplitude $\mathcal{A}$, due to the adoption of the Newtonian relation $\vesc(r) = \sqrt{-2\Phi_\mathrm{N}(r)}$, which neglects the effects of cosmic expansion and acceleration \citep[e.g.,][]{Miller2016,Rodriguez2024}. In this context, future extensions of the \clumiplus framework that incorporate these effects may help disentangle such contributions and refine the interpretation of phase-space observables in the infall regime.

\section{Summary and Conclusions}
\label{sec:summary}

We have presented \clumiplus, a self-consistent, multiprobe framework for reconstructing the mass distribution in and around galaxy clusters by combining gravitational lensing and dynamics (Sections~\ref{sec:lens} and \ref{sec:caustic}). In particular, \clumiplus highlights the power of combining redshift surveys with wide-field weak-lensing data to probe the cluster mass distribution. Building on the joint likelihood formalism of \citet{Umetsu2013}, \clumiplus integrates lensing profiles with projected escape velocity measurements in cluster outer regions. This approach enables constraints on the gravitational potential over a wide radial range---from the cluster core to the infall regions---without relying on equilibrium assumptions.

To achieve this broad radial coverage, \clumiplus incorporates strong-lensing aperture-mass constraints in the central region, weak-lensing shear and magnification profiles at larger radii, and projected escape velocity measurements in the outer infall regime (Section~\ref{sec:clumiplus}). The mass distribution is modeled using a flexible, piecewise-defined convergence profile within the lensing aperture, which transitions to a projected power-law form at larger radii. This hybrid prescription enables a self-consistent integration of observables across distinct physical regimes. Importantly, \clumiplus breaks both the mass-sheet degeneracy in lensing ($\lambda=1-\kappa_0$; Section~\ref{subsec:basics_lens}) and the velocity-scaling degeneracy in phase-space analysis ($\mathcal{G}$; Section~\ref{subsec:caustic}) by combining physically distinct tracers of the gravitational potential.

The \clumiplus framework is readily extensible to ensemble cluster analyses, enabling joint modeling of stacked weak-lensing and phase-space profiles for statistical samples of galaxy clusters. This capability makes it well suited for precision studies of outer halo structure, dynamical boundary features such as the splashback radius (Section~\ref{subsec:discussion}), and population-level trends in the mass distribution of clusters.

We validated \clumiplus using synthetic weak-lensing and phase-space observations of a massive cluster drawn from the IllustrisTNG simulation suite (Section~\ref{sec:tng}). The reconstructed mass profiles closely match the true simulation inputs across all three orthogonal projections, both in 2D and 3D, demonstrating robust performance even in the presence of triaxiality and projection effects. In the infall regions ($r_\perp>\Rinf=2\Mpch$), where escape velocity data are incorporated, we achieved precision gains of $10\percent$--$30\percent$ relative to lensing-only reconstructions (Section~\ref{subsec:gains}).

Within the radial regime where weak-lensing and phase-space constraints overlap, the precision gain increases with radius, suggesting that broader radial overlap between the two data sets will further enhance the statistical power of the joint analysis. This highlights the importance of coordinated spectroscopic and imaging surveys with overlapping coverage to fully exploit the capabilities of the \clumiplus framework in future wide-field applications.

In addition to improved statistical precision, we find that \clumiplus also delivers more accurate mass reconstructions (Section~\ref{subsec:accuracy}). Across all projections, \clumiplus consistently yields smaller fractional deviations from the ground truth compared to \clumi. Moreover, the reconstruction accuracy is better than the statistical precision, indicating that the inferred convergence profiles are statistically consistent with the true profile and show no evidence of systematic bias. These results confirm that incorporating escape velocity information enhances both the precision and fidelity of the mass reconstruction.

Furthermore, the framework demonstrates robustness against projection effects that can impact the identification of caustic boundaries, particularly those caused by chance alignments of line-of-sight substructures (Section~\ref{subsec:gains}). While such configurations may locally reduce statistical precision, they do not bias the inferred mass profiles. This robustness arises from the use of complementary, physically distinct probes of the gravitational potential, which respond differently to line-of-sight contamination. As a result, discrepancies between lensing and dynamical tracers can serve as diagnostics of complex line-of-sight structure rather than sources of systematic error.

As a first application to real data, we analyzed the galaxy cluster A2261 at $z=0.225$, combining Subaru and \HST weak+strong lensing data with HeCS spectroscopic measurements (Section~\ref{sec:a2261}). The reconstructed profile spans $r_\perp \in [30, 4000]\kpch$ and is consistent with previous CLASH lensing-only results, while achieving improved precision in the outer regions. These results highlight the utility of \clumiplus in systems where equilibrium assumptions may not apply.


In summary, \clumiplus provides a flexible, data-driven framework for mapping the mass distribution of galaxy clusters across a wide range of physical scales, achieving improved precision with minimal modeling assumptions. Its ability to integrate lensing and dynamical information makes \clumiplus particularly well suited for cosmological applications in the era of wide-field imaging and spectroscopic surveys---such as DESI \citep{DESI2016}, the Subaru Prime Focus Spectrograph \citep{Takada2014}, DESI II \citep{Schlegel2022}, LSST \citep{LSST2019}, \Euclid \citep{Euclid2025}, and the Roman High Latitude Survey \citep{Wang2022}---where accurate, unbiased mass calibration across large cluster samples is essential for precision cosmology \citep[][]{Tam2022}.

In future work, we plan to extend the \clumiplus framework to incorporate the effects of cosmic expansion acceleration in modeling escape velocities, thereby enhancing its applicability to accurate reconstructions of phase-space structure in the cluster outskirts and infall regions.


\begin{acknowledgements}
We thank the anonymous referee for providing insightful comments and suggestions. We thank Benedikt Diemer and Sandor Molnar for insightful discussions. This work is supported by the National Science and Technology Council, Taiwan (grant NSTC~112-2112-M-001-027-MY3) and by the Academia Sinica Investigator award (grant AS-IA-112-M04). M.P. acknowledges support from the Canada Research Chair Program and the Natural Sciences and Engineering Research Council of Canada (NSERC; funding reference No. RGPIN-2018-05425). A.D. acknowledges partial support from the INFN grant Indark. The Smithsonian Institution supports M.J.G.'s research. This work made use of high-level data products from the CLASH program \citep{Postman2012clash,Umetsu2016clash} and the Hectospec Cluster Survey \citep{Rines2013HeCS,pizzardo2020}, based on observations obtained with the Hubble Space Telescope, Subaru Telescope, and MMT Observatory. This research has also made use of NASA's Astrophysics Data System Bibliographic Services. \clumiplus is a proprietary code developed by the corresponding author in modern Intel FORTRAN.
\end{acknowledgements}


\facilities{Subaru (Suprime-Cam), HST (ACS, WFC3), MMT (Hectospec).}

\software{NumPy \citep{NumPy2020}, SciPy \citep{SciPy2020}, Matplotlib \citep{Hunter2007}, GetDist \citep{Lewis2019}.}

\FloatBarrier


\begin{appendix}

\section{Cluster Mass Model}
\label{appendix:model}

This appendix details the hybrid mass model implemented in \clumiplus (Section~\ref{subsec:model}), which combines a piecewise-defined convergence profile within the lensing aperture with an external power-law component to characterize the projected mass distribution at larger radii.

\subsection{Discretized Expressions for Cluster Lensing Profiles}
\label{subsec:estimators}

We derive discretized expressions for the lensing observables used in this work (Section~\ref{sec:lens}), based on the azimuthally averaged convergence profile $\kappa_\infty(\theta)$, modeled as a piecewise-constant function across radial bins.

Let $\{\theta_i\}_{i=1}^{N+1}$ denote a set of $(N+1)$ concentric radii spanning the full range $\theta \in  [\theta_1, \theta_{N+1}] \equiv [\thetamin,\thetamax]$, dividing the radial domain into $N = \NSL + \NWL$ annular bins. The average convergence within radius $\theta_i$ is then given by the discretized estimator:
\begin{equation}
 \label{eq:avkappa_d}
 \begin{aligned}
  \overline{\kappa}_\infty(<\theta_i) &= 
  \left(\frac{\thetamin}{\theta_i}\right)^2
  \overline{\kappa}_\infty(<\thetamin) \\
  &\quad + \frac{2}{\theta_i^2} \sum_{j=1}^{i-1}
  \Delta\ln\theta_j \,
  \overline\theta_j^2 \,
  \kappa_\infty(\overline\theta_j),
 \end{aligned}
\end{equation}
where $\Delta\ln\theta_j = (\theta_{j+1} - \theta_j)/\overline\theta_j$, and $\overline\theta_j$ denotes the area-weighted center of the $j$th bin in $[\theta_j, \theta_{j+1}]$.

In the strong-lensing regime, the constraints are provided in the form of aperture masses $\Map(<\theta_i)$ at $\NSL$ inner radii $\{\theta_i\}_{i=1}^{\NSL}$:
\begin{equation}
\label{eq:map}
  \Map(<\theta_i) = \pi (D_l \theta_i)^2 \SigmaCritInf \, \overline{\kappa}_\infty(<\theta_i),
\end{equation}
where $\SigmaCritInf = c^2 / (4\pi G D_l \beta_\infty)$ is the far-background critical surface mass density for lensing (Section~\ref{subsec:Nz}).

For the weak-lensing regime ($i = \NSL+1, \dots, N$), the observables are azimuthally averaged in annular bins centered at $\overline\theta_i$. The discretized forms of the tangential reduced shear and inverse magnification are
\begin{equation}
 \begin{aligned}
 \langle g_+ \rangle(\overline\theta_i) &=
  \frac{
  \langle W \rangle_g \left[
  \overline{\kappa}_\infty(<\overline\theta_i) - \kappa_\infty(\overline\theta_i)
  \right]
  }{
  1 - f_g \langle W \rangle_g \kappa_\infty(\overline\theta_i)
  },\\
  \langle \mu^{-1} \rangle(\overline\theta_i) &=
  \left[1 - \langle W \rangle_\mu \kappa_\infty(\overline\theta_i) \right]^2\\
  &\quad - \langle W \rangle_\mu^2 \left[
  \overline{\kappa}_\infty(<\overline\theta_i) - \kappa_\infty(\overline\theta_i)
  \right]^2.
 \end{aligned}
\end{equation}

Under the piecewise-constant assumption, the interior average $\overline{\kappa}_\infty(<\overline\theta_i)$ at the center of the $i$th radial bin is given by
\begin{equation}
\begin{aligned}
  \overline{\kappa}_\infty(<\overline\theta_i) =
 \frac{1}{2}
 \Bigg[
 &\left( \frac{\theta_i}{\overline\theta_i} \right)^2 
 \overline{\kappa}_\infty(<\theta_i) \\
 + 
 &\left( \frac{\theta_{i+1}}{\overline\theta_i} \right)^2 \overline{\kappa}_\infty(<\theta_{i+1})
 \Bigg],
\end{aligned}
\end{equation}
with $\overline{\kappa}_\infty(<\theta_i)$ and $\overline{\kappa}_\infty(<\theta_{i+1})$ computed using Equation~(\ref{eq:avkappa_d}).

All lensing observables in \clumiplus (Section~\ref{sec:lens}) are thus uniquely determined by a set of $N+1$ piecewise-defined convergence parameters:
\begin{equation}
\bs = \left\{ \kappa_{\infty,\mathrm{min}}, \kappa_{\infty,i} \right\}_{i=1}^N,
\end{equation}
where $\kappa_{\infty,\mathrm{min}} \equiv \kappa_\infty(<\thetamin)$ represents the central convergence interior to $\thetamin$, and $\kappa_{\infty,i} \equiv \kappa_\infty(\overline\theta_i)$ denotes the convergence in each radial bin.

\subsection{Predicting the Escape Velocity}
\label{appendix:pred_vesc}

Including the external convergence component defined in Equation~(\ref{eq:kext}), the full radial convergence profile $\kappa_\infty(\theta)$ can be expressed as
\begin{equation}
 \label{eq:kappa_theta_full}
 \begin{aligned}
 \kappa_\infty(\theta) &= \overline{\kappa}_{\infty,\mathrm{min}}\, \Theta(\thetamin - \theta)\\
 &\quad + \sum_{i=1}^{N} \kappa_{\infty,i} \, \Theta(\theta - \theta_i, \theta_{i+1} - \theta)\\
 &\quad + \kappa_{\infty,\mathrm{ext}}\, \Theta(\theta - \thetamax)   \left( \frac{\theta}{\thetamax} \right)^{-q},
 \end{aligned}
\end{equation}
where $\Theta(x)$ denotes the Heaviside step function and $\Theta(x, y) \equiv \Theta(x)\Theta(y)$ restricts support to the interval $[\theta_i, \theta_{i+1})$.
The first term represents the average convergence within the innermost aperture ($\theta < \thetamin$), parameterized by $\overline{\kappa}_{\infty,\mathrm{min}}$. The second term describes the piecewise-constant convergence profile over the lensing annuli $\theta \in [\thetamin, \thetamax]$, with bin-averaged values $\kappa_{\infty,i}$. The final term models the projected mass distribution outside the lensing aperture ($\theta \ge \thetamax$) as a power-law profile with normalization $\kappa_{\infty,\mathrm{ext}}$ and logarithmic slope $q$.

By construction, the external convergence term contributes only for $\theta > \thetamax$ and does not affect lensing observables within the lensing field. To constrain this outer component, complementary information beyond lensing is required. Within the \clumiplus framework, projected escape velocity measurements at $r_\perp > \Rinf$ provide such constraints by probing the gravitational potential in the cluster outskirts.

For $\theta \ge \thetamax$, the average convergence interior to $\theta$ is given by
\begin{equation}
 \label{eq:avg_kappa_ext}
 \begin{aligned}
  \overline{\kappa}_\infty(<\theta) &= 
  \left( \frac{\thetamax}{\theta} \right)^2 \Bigg[ 
  \overline{\kappa}_\infty(<\thetamax) \\
  &\quad + \frac{2\kappa_{\infty,\mathrm{ext}}}{2 - q} 
  \left( \left( \frac{\theta}{\thetamax} \right)^{2 - q} - 1 \right)
  \Bigg],
 \end{aligned}
\end{equation}
where we assume $0 < q < 2$ to ensure convergence of the potential integral. The corresponding projected aperture mass, $\Map(<\theta)$, is computed using Equation~(\ref{eq:map}).

The spherically enclosed mass profile $M(<r)$ is obtained via the Abel integral transform (Equation~(\ref{eq:abel})), and the Newtonian gravitational potential $\Phi_\mathrm{N}(r)$ is then computed from Equation~(\ref{eq:Poisson}).

Given a set of model parameters $\bm = \{\bs, \mathcal{G}, \kappa_{\infty,\mathrm{ext}}, q\}$, the projected escape velocity profile as a function of $r_\perp = D_l \theta$ is given by
\begin{equation}
 \mathcal{A}(r_\perp) = \sqrt{\frac{-2\Phi_\mathrm{N}(r_\perp)}{\mathcal{G}}},
\end{equation}
where $\mathcal{G}$ is the depletion factor relating the Newtonian escape velocity to the observable caustic amplitude (Section~\ref{sec:caustic}).

\begin{figure*}[htb] 
 \begin{center}
  \includegraphics[width=0.33\textwidth,angle=0,clip]{\FIG/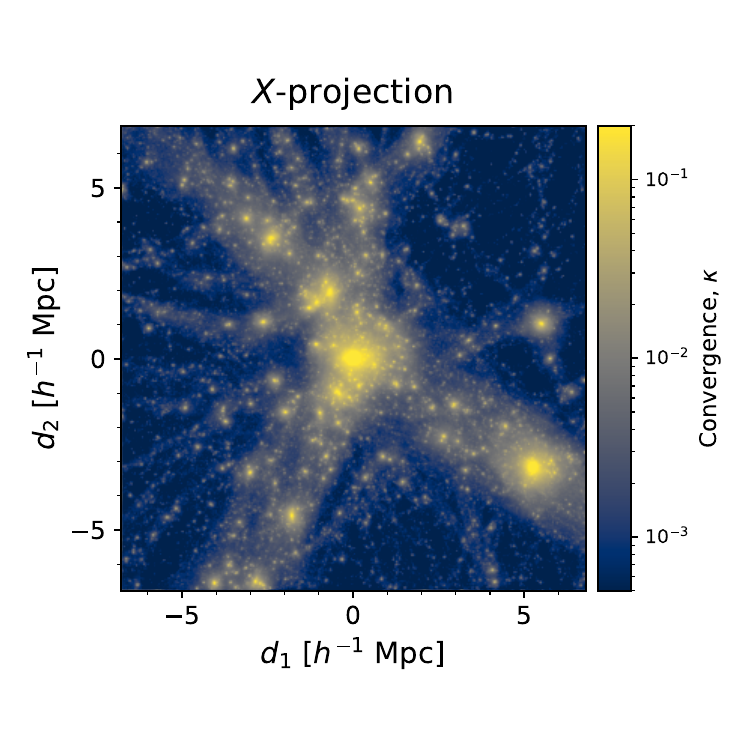}
  \includegraphics[width=0.33\textwidth,angle=0,clip]{\FIG/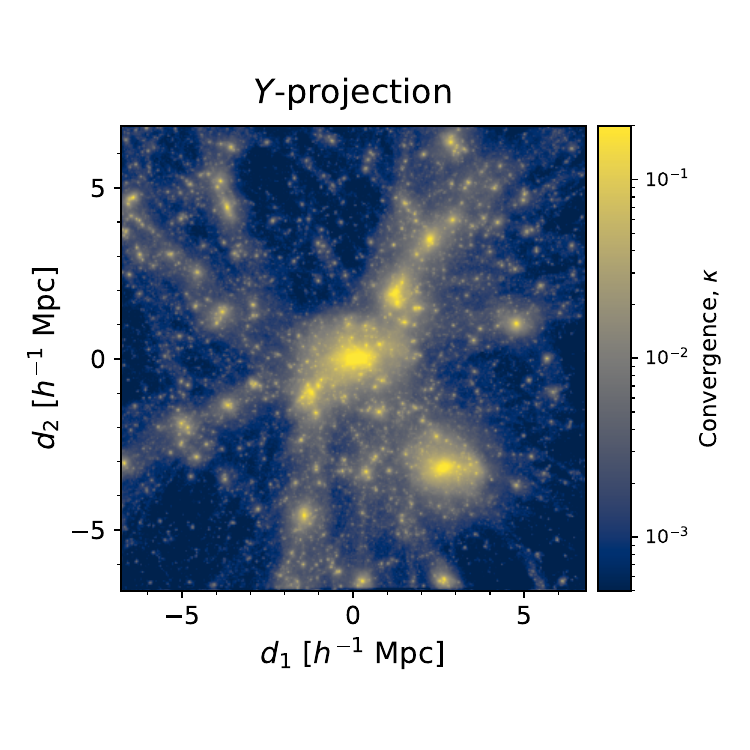}
  \includegraphics[width=0.33\textwidth,angle=0,clip]{\FIG/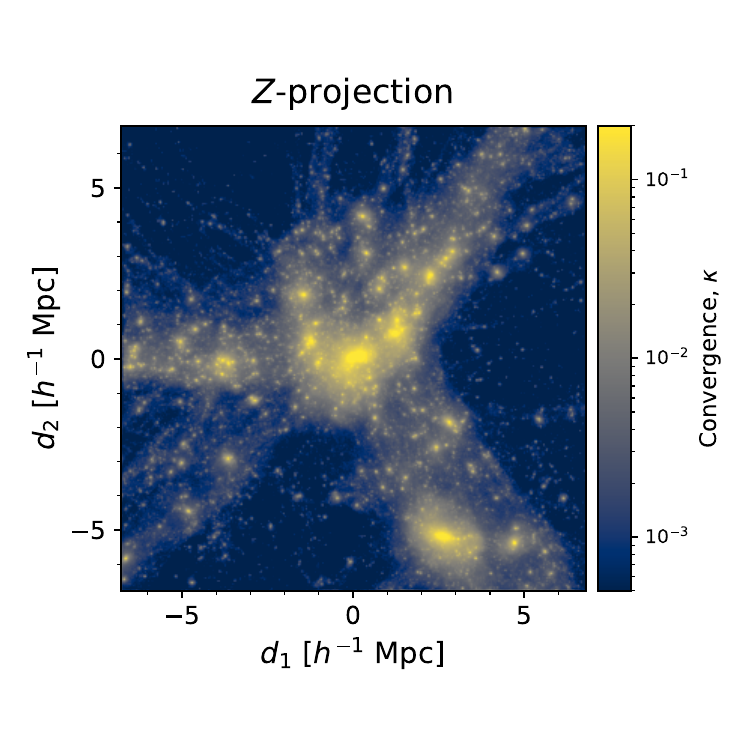}
 \end{center}
\caption{Projected mass maps centered on a massive galaxy cluster at $z_\mathrm{cl} = 0.21$ from the IllustrisTNG-300-1 simulation, shown along three orthogonal directions: $X$ (left), $Y$ (middle), and $Z$ (right) axes. The color bar indicates the lensing convergence, $\kappa = \Sigma / \SigmaCrit$, for background sources located at $\zs = 1.2$.
\label{fig:tng_maps}}
\end{figure*}

\begin{figure*}[htb] 
 \begin{center}
  \includegraphics[width=0.33\textwidth,angle=0,clip]{\FIG/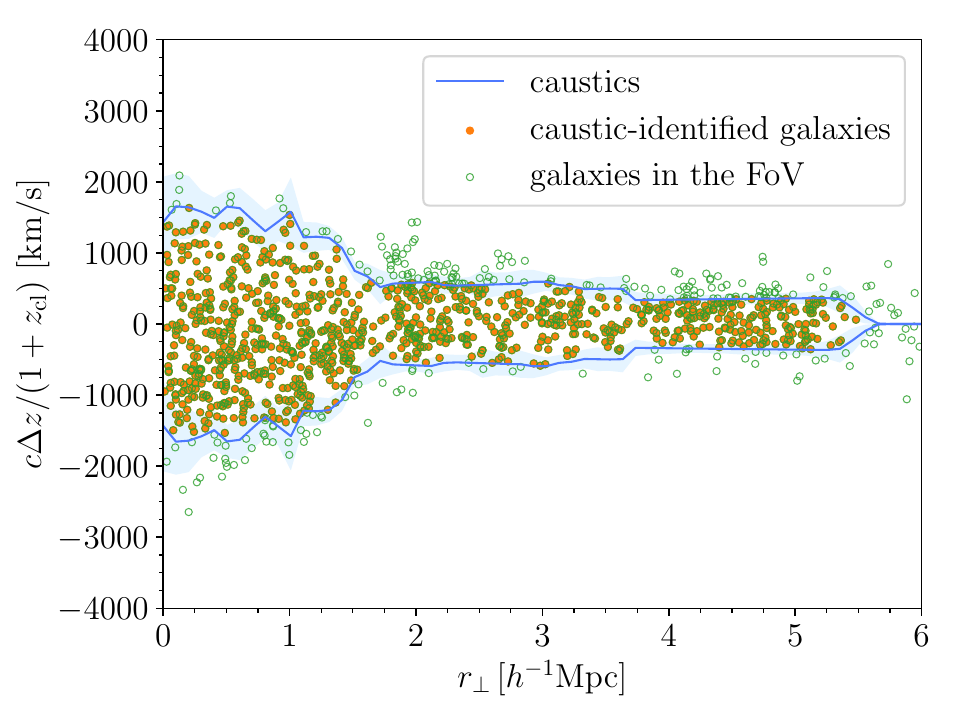}
  \includegraphics[width=0.33\textwidth,angle=0,clip]{\FIG/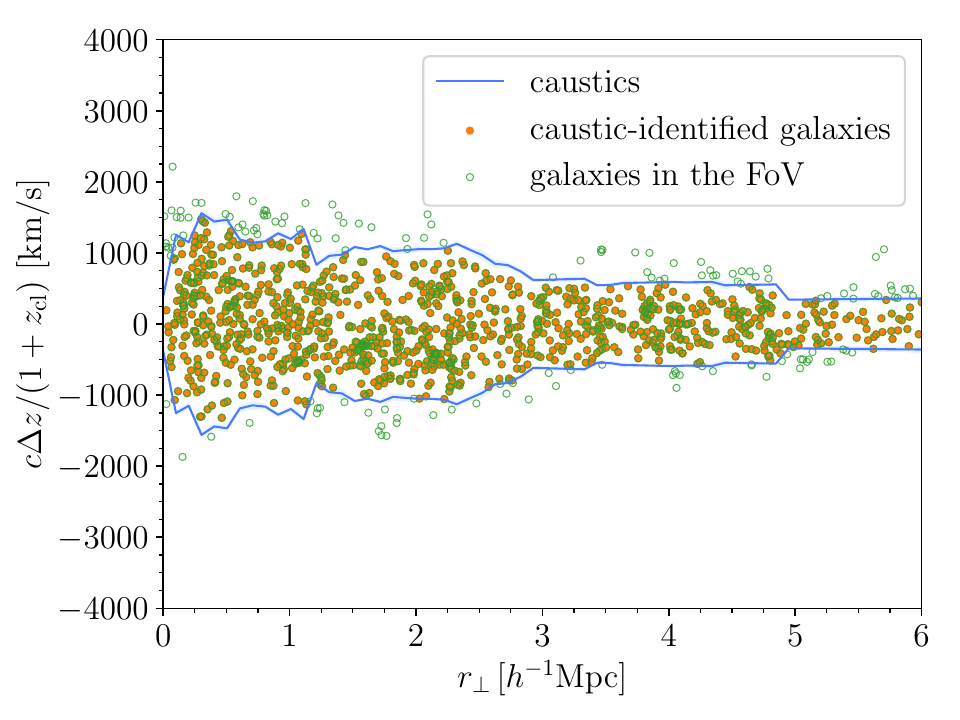}
  \includegraphics[width=0.33\textwidth,angle=0,clip]{\FIG/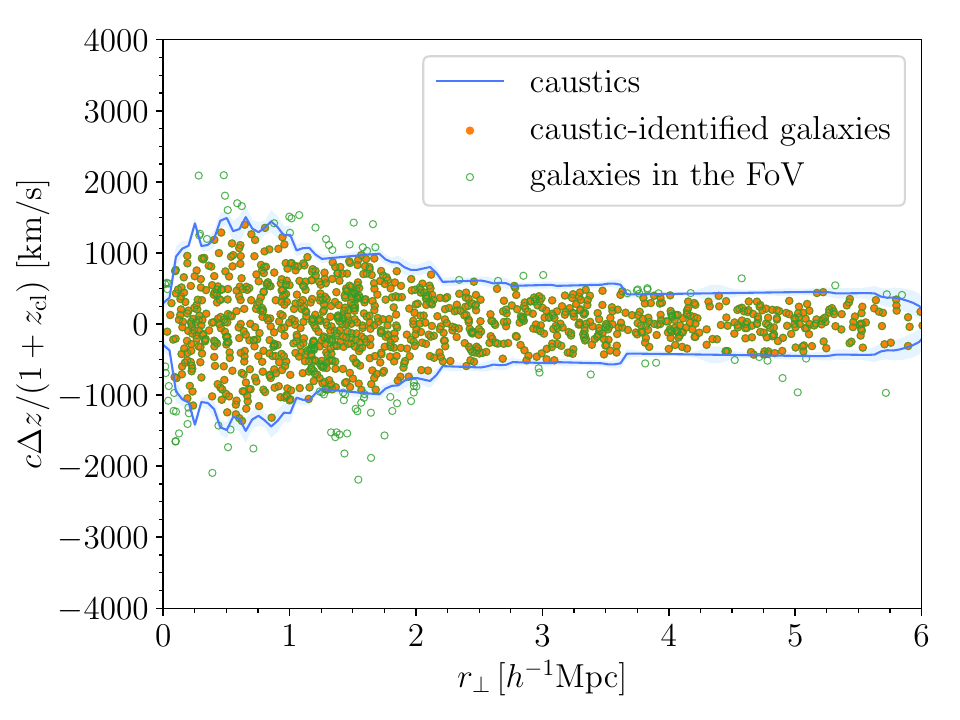}
 \end{center}
\caption{Redshift diagrams of the TNG300-1 cluster at $z_\mathrm{cl} = 0.21$ (corresponding to the projected mass maps in Figure~\ref{fig:tng_maps}), shown along three orthogonal projections: $X$ (left), $Y$ (middle), and $Z$ (right) axes. Each panel shows the line-of-sight velocity of galaxies, $v_\mathrm{los}=c\Delta z/(1+z_\mathrm{cl})$, relative to the BCG, plotted as a function of projected cluster-centric radius, $r_\perp$. Green circles denote all galaxies within the field of view, while filled orange circles represent galaxies located within the caustic boundaries, corresponding to candidate bound or infalling systems.
\label{fig:tng_caustic}}
\end{figure*}

\section{Synthetic Data from the IllustrisTNG 300-1 Simulation}
\label{appendix:tng}

This appendix presents the projected mass maps and redshift diagrams used to construct the synthetic weak-lensing and phase-space observables for the TNG300-1 cluster at $z = 0.21$, as described in Section~\ref{sec:tng}. Figure~\ref{fig:tng_maps} displays the projected mass maps $\kappa = \Sigma/\SigmaCrit$ for sources at $\zs = 1.2$, shown for three orthogonal projections ($X$, $Y$, and $Z$), each centered on the BCG.  These maps are used to generate the azimuthally averaged lensing profiles shown in Figure~\ref{fig:tng_lens}.

Figure~\ref{fig:tng_caustic} presents the corresponding redshift diagrams, showing line-of-sight velocity versus projected cluster-centric radius for the same three projections. These phase-space diagrams reveal the caustic structures from which the projected escape velocity profiles $\mathcal{A}(r_\perp)$ are extracted using the method described in Section~\ref{subsec:tng_caustic}. Notably, the $Y$-projection exhibits a broadened and elevated caustic envelope, resulting from two infalling substructures aligned along the line of sight. This feature highlights the sensitivity of phase-space observables to projection effects.

\section{Observational Profiles for A2261}
\label{appendix:a2261_data}

This appendix provides the input weak-lensing and projected escape velocity profiles used in the joint \clumiplus analysis of A2261 at $z=0.225$ (Section~\ref{sec:a2261}). These figures visualize the observational constraints derived from Subaru imaging and HeCS spectroscopy, as described in Section~\ref{subsec:a2261_data}.

\begin{figure}[htb!] 
 \begin{center}
  \includegraphics[width=0.9\columnwidth,angle=0,clip]{\FIG/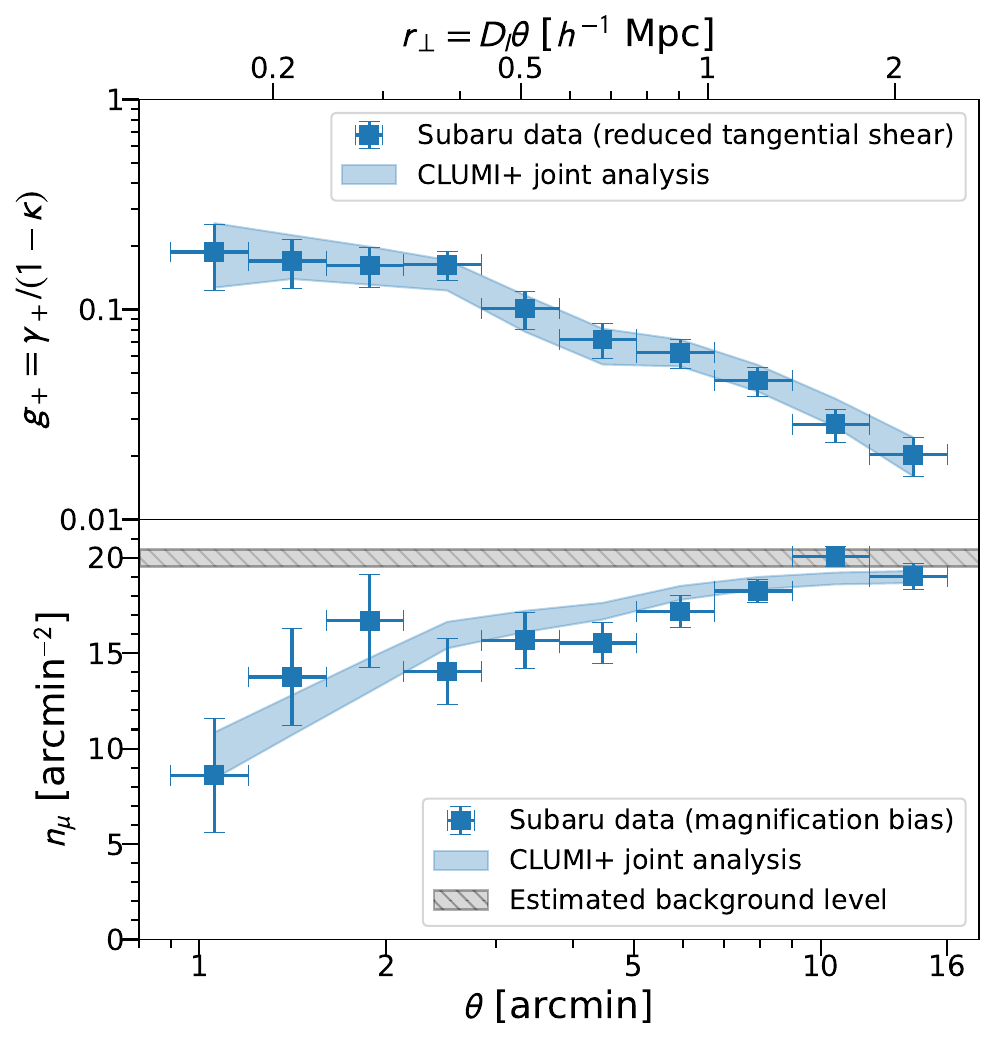}
 \end{center}
\caption{
Azimuthally averaged weak-lensing profiles of A2261 derived from Subaru observations. The top panel shows the reduced tangential shear profile, $\langle g_+\rangle(\theta)$ (squares), while the bottom panel presents the magnified source number counts of background galaxies, $\langle n_\mu\rangle(\theta)$ (squares). In each panel, the shaded region indicates the marginalized $1\sigma$ confidence interval from the joint \clumiplus analysis. The horizontal bar (gray shaded and hatched region) denotes the estimated mean background level.
\label{fig:a2261_wl}
}
\end{figure}

\begin{figure}[htb!] 
 \begin{center}
  \includegraphics[width=0.9\columnwidth,angle=0,clip]{\FIG/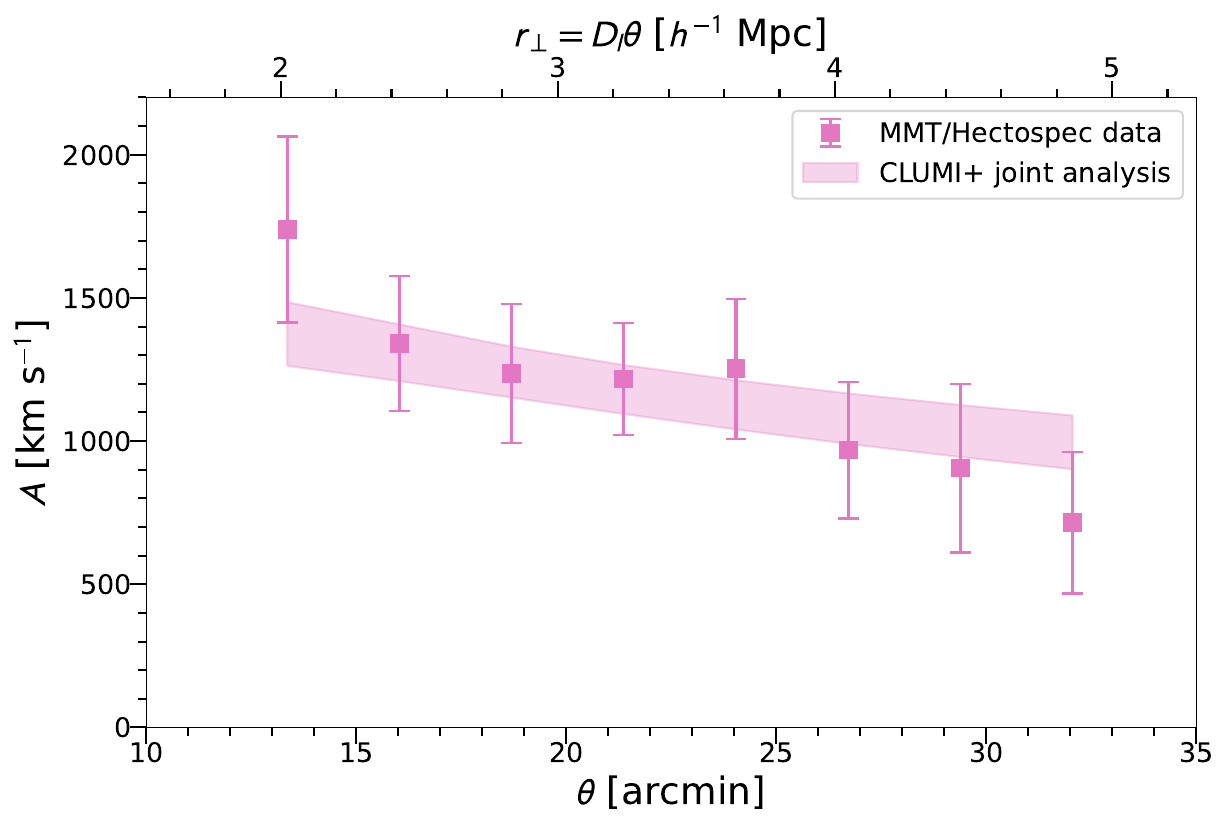}
 \end{center}
\caption{Projected escape velocity profile of A2261 derived from HeCS spectroscopic observations (Figure~\ref{fig:a2261_rvlos}). Squares denote the measured caustic amplitudes, with uncertainties that include both observational errors and an additional 20\percent scatter to account for projection effects. The shaded region shows the marginalized $1\sigma$ confidence interval from the joint \clumiplus analysis. 
\label{fig:a2261_vesc}}
\end{figure}

Figure~\ref{fig:a2261_wl} shows the Subaru weak-lensing measurements of the tangential distortion and magnification bias profiles. Figure~\ref{fig:a2261_vesc} displays the projected escape velocity profile $\mathcal{A}(r_\perp)$ at $r_\perp > 2\Mpch$ derived from spectroscopic galaxy redshifts using the caustic technique (see Figure~\ref{fig:a2261_rvlos}).

\end{appendix}




\end{document}